\documentclass[%
 reprint,
 amsmath,amssymb,
 aps,
]{revtex4-2}

\usepackage{graphicx}
\usepackage{dcolumn}
\usepackage{bm}
\usepackage{epstopdf}
\usepackage{caption}
\usepackage{subcaption}
\usepackage{braket}
\usepackage{xcolor}
\usepackage{overpic}
\usepackage{hyperref}

\newcommand\sbullet[1][.5]{\mathbin{\vcenter{\hbox{\scalebox{#1}{$\bullet$}}}}}
\usepackage{csquotes}

\begin{document}

\preprint{APS/123-QED}

\title{Inertial Particle Dynamics in Traveling 
Wave Flow}

\author{P. Swaathi}
 \email{swaathi.p2019@vitstudent.ac.in}

\author{Sanjit Das}%
 \email{sanjit.das@vit.ac.in}
\affiliation{%
Division of Physics\\
 School of Advanced Sciences\\
 Vellore Institute of Technology, Chennai, Tamil Nadu - 600 127, India.
}%

\author{N. Nirmal Thyagu}
 \email{nirmalthyagu@mcc.edu.in}
\affiliation{
 Department of Physics,\\
 Madras Christian College (Autonomous)\\
 [Affiliated to the University of Madras],\\
 Chennai, Tamil Nadu - 600 059, India.
}%

\date{\today}

\begin{abstract}
The dynamics of inertial particles in fluid flows have been the focus of extensive research due 
to their relevance in a wide range of industrial and environmental processes. The Maxey-Riley 
equation and its simplified versions are widely employed to study the behavior of inertial 
particles and they demonstrate extremely rich dynamics. In particular, the clustering and 
segregation phenomena exhibited by aerosols and bubbles in fluid flows are quite non-trivial. 
Earlier studies have examined the dynamics of aerosols and bubbles using a simplified Maxey-Riley equation
in some standard systems that include Karmon-vortex flow, double-gyre flow and also examined the dynamics of tracer particles in cellular flow and shear flow systems. But the dynamics within traveling wave flows remain unexplored. The traveling wave flow is a simple two-dimensional incompressible fluid flow that exhibits both homoclinic and hetroclinic trajectories. In this paper, we investigated the dynamics of inertial particles in a traveling wave flow using the simplified Maxey-Riley equation. The inertial particle flow displays rich dynamics, mixing, and segregation in phase space as well as the emergence of Lagrangian Coherent Structures (LCS). We first obtain the finite-time Lyapunov exponent (FTLEs) for the base fluid flow defined by the traveling wave flow using the Cauchy-Green deformation tensor. Further, we extend our calculations to the inertial particles to get the inertial finite-time Lyapunov exponent (iFTLEs). Our findings reveal that heavier inertial particles tend to be attracted to the ridges of the FTLE fields, while lighter particles are repelled. By understanding how material elements in a flow separate and stretch, one can predict pollutant dispersion, optimize the mixing process, and improve navigation and tracking in fluid environments. This provides insights into the complex and non-intuitive behavior of inertial particles in chaotic fluid flows, and may have implications for pollutant transport in wide-ranging fields such as atmospheric and oceanic sciences.

\begin{description}
\item[Keywords]
Lagrangian coherent structures, inertial particles, finite-time Lyapunov exponent. 
\end{description}
\end{abstract}

\maketitle


\section{\label{sec:level1}Introduction}

The study of motion of rigid particles in fluids has been of interest since the work of Poisson \cite{poisson}. Inertial particle flows hold a significant role in both natural and industrial settings, necessitating a deeper understanding of their dynamics.  The equation of motion describing the inertial particles was given by Maxey and Riley in 1983 known as the Maxey-Riley equation \cite{maxey1983equation}.
Following this, a large number of studies have looked into how inertial particles disperse within flows, particularly in turbulent settings
 \cite{ bec2007heavy, squires1991preferential, riley1974diffusion, bec2003fractal, bec2005clustering}. This dynamic behavior of particles extends to phenomena like gravitational settling velocity and settling time \cite{maxey1986gravitational, maxey1987motion, maxey1987gravitational, rubin1995settling}. Notably, the phenomenon of preferential concentration of inertial particles within specific regions of fluid flow has significant attention due to its relevance in various applications, such as particle dispersion by clouds \cite{shaw1998preferential}, oil spill dynamics \cite{mezic2010new}, urban pollution studies \cite{tang2012geometry}, oceanic studies \cite{beron2008oceanic, nencioli2011surface} and ecological investigations like plankton dynamics \cite{espinosa2015density, peng2009transport}.

In this paper, we study the phenomena of preferential concentration of inertial particles and how it is influenced by the finite-time Lyapunov exponent (FTLE) fields of the underlying fluid. The ridges of FTLEs form the Lagrangian Coherent Structures (LCS) which are the spatio-temporal structures that serve as barriers to transport. We also extended our study to find the inertial FTLE (iFTLE) fields, which is derived from measuring the separation of inertial particles in the flow.  Extending previous research \cite{sapsis2009inertial, sapsis2008instabilities, haller2008inertial} we simulate the Lagrangian trajectories of particles lighter than the ambient fluid (bubbles) and those heavier than the ambient fluid (aerosols). This approach provides a broader perspective on inertial particle dynamics and their preferential concentration effects.

We have chosen the base fluid flow to be the traveling wave flow which models a two-dimensional steady and incompressible flow, and has common features that can be identified with many fluid systems. This flow was previously analyzed by Weiss \cite{weiss1991transport, weiss1989mass} who showed that it has interesting transport and mixing properties that are relevant to phenomena such as acid rain and ozone depletion. However, the dynamics of inertial particles in the traveling wave flow have not been explored yet. The traveling wave flow system exhibits the co-existence of homoclinic and heteroclinic trajectories and the effects of inertial particles are not yet studied for this particular system which explicitly alters the transport and mixing behavior. In this paper, we explore how the inertial particles behave in the traveling wave flow, and how their dynamics differ from the fluid particles. We also examine how the inertial particles affect the dispersion, clustering, and mixing properties in the flow.Understanding the dispersion and accumulation of inertial particles holds significant potential applications in turbulent flows, particularly in pollution control, cloud formation, and industrial mixing \cite{fung1998two, chen2006turbulent, biferale2005lagrangian, batchelor1952diffusion, boffetta2002statistics}.

In this study, our primary goal is to study the patterns that emerge as we vary the Stokes number ($St$) \cite{tallapragada2008particle} and the density ratio ($R$) of inertial particles. The Stokes number reflects the nondimensional particle response time relative to the hydrodynamic time scale of the flow. We observe how inertial particles accumulate along thin flow structures, influenced by the parameters $St$ and $R$. Notably, aerosol particles ($R$ = 0) with lower Stokes numbers behave more akin to incompressible fluid particles, while bubble-like particles ($R$ = 1) exhibit similar patterns as $St$ increases. Dynamical systems approaches have been successful in understanding the behavior of passive fluid particles, particularly through the identification of stable and unstable manifolds in fluid flows, which reveal key separatrices. Traditionally, the maximum Lyapunov exponent has been used to quantify the rate of separation of nearby trajectories in a flow. However, this measure often falls short when dealing with time-dependent and spatially varying flows because it captures the average separation over a long time and overlooks transient and localized effects. Whereas, finite-time Lyapunov exponent (FTLE) fields provide a more nuanced and computationally tractable method for studying the behavior of inertial particles. FTLE fields are calculated using the Cauchy-Green deformation tensor and they identify regions of attraction or repulsion in the flow. This offers an effective computationally tractable method to study the behavior of inertial particles \cite{mehlig2005aggregation}. Past studies have employed FTLE for various scenarios, such as airborne microbes \cite{tallapragada2011lagrangian}, urban flows \cite{tang2012geometry, tang2009locating}, and turbulent boundary layers, to gain insights into particle dynamics \cite{peng2009transport, sapsis2009inertial, haller2008inertial, beron2015dissipative}.

Our work builds on this foundation by systematically investigating the dynamics of inertial particles using FTLE fields \cite{perez2015clustering, babiano2000dynamics}. We consider the behavior of particles with varying density ratios, in addition to varying Stokes numbers. We find that the dynamics of aerosols are influenced by the Stokes number, leading to their attraction to FTLE ridges. In contrast, bubbles exhibit a repulsion, driven by their density and Stokes number. FTLE fields provide information about attractor and repeller structures within the flow, however, they fall short when it comes to understanding the dynamics of inertial particles. FTLE fields are used to analyze the behavior of neutrally buoyant fluid particles, which can be limiting in capturing the complex mixing behaviors of inertial particles. Whereas, inertial finite-time Lyapunov exponent (iFTLE) fields quantify the mixing and dispersion of inertial particles which will be useful for modeling and predicting the dynamics of such particles in various industrial processes. It also provides insights into the complex interactions between inertial particles and fluid flows.

The paper is organized as follows. In section \ref{sec: second} we provide the background and methodology for calculating inertial particle dynamics using the Maxey-Riley equation and the finite-time Lyapunov exponent. Following this, in section \ref{sec: third} we discuss the traveling wave flow and its properties by looking at the Eulerian fields. Further, we study the Lagrangian evolution of fluid tracers and inertial particles system and investigate the behavior of these particles by varying their Stokes number, density ratio, and integration times. In section [\ref{sec: fourth}] we conclude our study and suggest some directions for future research.

\section{Background Methodology}\label{sec: second}

In this section, we provide a short background of the Maxey-Riley equation \cite{maxey1983equation} that describes the dynamics of finite-sized inertial particles in fluid flows. Following this we discuss the method to calculate the finite-time Lyapunov exponents for fluid particles using the Cauchy-Green deformation tensor. And this method is also extended to the inertial particles for the calculation of its corresponding exponents called as inertial finite-time Lyapunov exponents (iFTLEs).

\subsubsection*{\textbf{A. Maxey-Riley Equation}}
Inertial particles have a different density from that of the surrounding fluid in which they are introduced. This contrasts with tracer particles, which have the same density as that of the fluid flow.  When numerically simulating the behavior of inertial particles, we need to integrate both their position and velocity. We use the simplified Maxey-Riley equation \cite{maxey1983equation} for our study. In its dimensional form, the equation is expressed as follows:

\begin{equation}\label{eq:1}
\begin{split}
m_p \dot{v} = m_f \frac{D}{Dt} u(r(t),t) + (m_p - m_f)g \\ 
-6\pi a \mu A(t) \\ 
- \frac{1}{2} m_f \frac{d}{dt} B(t) \\
- 6 \pi a^2 \mu \int_0^t d\tau \frac{\frac{dX(t)}{d\tau}}{\sqrt{\pi \nu (t - \tau)}}
\end{split}
\end{equation}

where
\begin{equation*}
A(t) = v(t) - u(r(t),t) - \frac{1}{6} a^2 \Delta^2 u
\end{equation*}

\begin{equation*}
B(t) = v -u(r(t),t)- \frac{1}{10} a^2\Delta ^2 u
\end{equation*}

Here, $m_p$ represents the mass of the inertial particle, while $m_f$ represents the mass of the fluid. The fluid's velocity, denoted as $u(r(t), t)$, is a function of both time ($t$) and the position of a particle ($r(t)$). The particle's velocity, v(t) = $\dot{r}(t)$, is defined as the velocity of the particle over time $t$. Additionally, the fluid's viscosity is represented by $\mu$, $\textit{a}$ signifies the particle's radius, and $\textit{g}$ corresponds to the acceleration due to gravity. The term $Du/Dt$ represents the total derivative, taken along a fluid element's path, and is expressed as $Du/Dt = \partial u/\partial t +(u \sbullet \Delta)u$.  The quantity $du/dt$, on the other hand, represents the total derivative, calculated along the trajectories of a particle, and is defined as $du/dt = \partial u/\partial t +(v \sbullet \Delta)u$. In Eq. (\ref{eq:1}), there are several distinct terms, each contributing to a different aspect of the particle's behavior in the fluid flow. The first term quantifies the force exerted on the particle by the surrounding undisturbed fluid. The second term accounts for buoyancy effects, considering the particle's weight in the fluid. The third term is associated with Stokes drag, reflecting the resistance encountered by the particle as it moves through the fluid. The fourth term incorporates added mass effects, which characterize the additional inertia experienced by the particle due to its interaction with the fluid. The fifth term, an integral component, is the Basset history term, known for its local modification of flow gradients. Additionally, the Faxén correction term is expressed as $a^2 \Delta^2 u$. The  Eq. (\ref{eq:1}) holds true for small, rigid, spherical particles at low Reynolds numbers. When the particle radius is sufficiently small, we can neglect the Faxén correction terms. Moreover, if the time it takes for a particle to revisit a previously explored region significantly exceeds the relevant time scale of the problem, we can safely neglect the Basset history term \cite{tel2005chemical, michaelides1997transient}.  Under these assumptions, Eq. (\ref{eq:1}) can be written as follows which is non-dimensionalized using the length scale (L) and the velocity scale (U),

\begin{equation}\label{eq:2}
\ddot{r}(t) = \frac{1}{St} \Bigr[u(r(t),t) - \dot{r}(t)\Bigr]- Hn + \frac{3}{2}R\frac{d}{dt}u(r(t),t).
\end{equation}

Here,
\begin{equation*}
\frac{1}{St} = \frac{6\pi a \mu L}{(m_p + \frac{1}{2}m_f)U}
\end{equation*}

\begin{equation*}
R = \frac{m_f}{m_p + \frac{1}{2}m_f}
\end{equation*}

\begin{equation*}
H = \frac{m_p - m_f}{6 \pi a \mu U St}g
\end{equation*}

In this study, gravity (\textit{g}) is not taken into account and \textit{n} is the unit pointing vector in the direction of gravity, therefore Eq. (\ref{eq:2}) becomes,

\begin{equation}\label{eq:3}
\ddot{r}(t) = \frac{1}{St}\Bigr[u(r(t),t) - \dot{r}(t) \Bigr] + \frac{3}{2}R\frac{d}{dt}u(r(t),t)
\end{equation}

In the above equation, $St$ represents the Stokes number, where $R$ represents the density ratio parameter between the particle and the surrounding fluid. The Stokes number ($St$), quantifies the relationship between a particle's characteristic time scale and that of the fluid flow. In cases of low Stokes numbers, particles tend to follow the same path as the fluid. Conversely, when Stokes numbers are high, particles exhibit distinct dynamics from that of the fluid flow. Coming to the density ratio parameter, when $R$ equals 2/3, it signifies that the particle has the same density as the fluid. If $R$ exceeds 2/3, the particles are lighter than the fluid, which typically corresponds to bubbles. Conversely, for $R$ values below 2/3, the particles are denser than the fluid, often characterizing aerosols.

\subsubsection*{\textbf{B. Finite-time Lyapunov Exponents (FTLEs)}}

In this section, we examined how the finite-time Lyapunov exponent is computed. This approach serves the purpose of determining the highest rate of expansion exhibited by neighboring particles within the fluid flow \cite{schuster2012transport, haller2001distinguished, shadden2005definition}.

Let's consider a scenario: at the starting moment $t_0$ = 0, a particle is situated at the position $x$. As time progresses to $t$, the particles undergo advection resulting in a new position denoted as $\Phi_{t_0}^{t_0+t}(x)$.

Now, let's introduce a perturbed point, initially positioned at $y = x+\Delta x(0)$ at time $t_0$. Following the passage of time $t$, this perturbation evolves

\begin{equation*}
\Delta x(t) = \Phi_{t_0}^{t_0+t}(y) - \Phi_{t_0}^{t_0+t}(x)
\end{equation*}

\begin{equation}\label{eq:4}
\Phi_{t_0}^{t_0+t}(y) - \Phi_{t_0}^{t_0+t}(x) = \frac{d\Phi_{t_0}^{t_0+t}(x)}{dx} \Delta x(0) + \mathcal{O}(\parallel \Delta x (0) \parallel^2)
\end{equation}

In these equations, $\Delta x(t)$ represents the displacement between the particles at different positions and times. The operator $\Phi_{t_0}^{t_0+t}$ represents the mapping of a particle's position from time ($t_0$) to ($t_0+t$). The term $d\Phi_{t_0}^{t_0+t}(x)/{dx}$ corresponds to the deformation gradient tensor. We simplify the expressions by neglecting the higher order terms in Eq. (\ref{eq:4}), then the magnitude of the perturbation is

\begin{equation*}
\parallel \Delta x(t)\parallel = \sqrt{\braket{\Delta x(0), \Lambda \Delta x(0)}}.
\end{equation*}

Here, 

\begin{equation*}
\Lambda = \frac{d\Phi_{t_0}^{t_0+t}(x)^{*}}{dx} \quad \frac{d\Phi_{t_0}^{t_0+t}(x)}{dx} 
\end{equation*}

is the Cauchy-Green deformation tensor, where the superscript ``$\ast$'' indicates the transpose of the tensor. Our objective is to determine the direction along which most extensive stretching happens and to measure its magnitude which is the maximum eigenvalue of the Cauchy-Green Tensor. That is, the greatest degree of stretching is achieved when $\Delta x(0)$ is oriented along the eigenvector corresponding to the largest eigenvalue of $\Lambda$, which is given by
\begin{equation}
\underset{\Delta x(0)}{\text{max}} \parallel \Delta x(t)\parallel = \sqrt{\lambda_{max}(\Lambda)}\parallel \overline{\Delta x}(0) \parallel .
\end{equation}

Then, the finite-time Lyapunov exponent (FTLE) is given by,

\begin{equation}
\sigma_{t_0}^{t_0 + t} (x) = \frac{1}{|t|} ln \sqrt{\lambda_{max} (\Lambda)},
\end{equation}

and it enables the integration over finite times ``$t$" that encompass both positive and negative values. Negative integration (backward-time integration) yields attracting Lagrangian coherent structures which reveal the unstable manifolds for a time-independent vector field, while positive integration (forward-time integration) yields repelling Lagrangian coherent structures which reveal the stable manifolds for a time-independent vector field. Essentially, for a given FTLE field, Lagrangian coherent structures (LCS) are defined as \textit{ridges} of the field.

\section{Particle clustering and segregation in fluid flow}\label{sec: third}

In this section, we start our study with the well-known Travelling Wave Flow (TWF) and briefly outline its salient features by depicting its Eulerian velocity field. We will use the traveling wave flows as the base fluid flow for investigating the dynamics of inertial particles. The TWF is a steady and an incompressible fluid flow that exhibits rich dynamical structures and complex patterns making it a good model for understanding how inertial particles behave in such flows. Next, we explore the Lagrangian trajectories of both fluid tracers and inertial particles. To visualize and analyze the flow field and to determine the barriers of transport, we use Finite-Time Lyapunov Exponent (FTLE) plots. Calculating the FTLE requires integrating the particle trajectories.  For fluid tracers, one has to integrate the velocity field using Eq. (\ref{eq:7}), whereas for inertial particles one has to integrate the position and velocity of the particle according to Eq. (\ref{eq:3}).

\subsubsection*{\textbf{A. Traveling Wave Flow}}

The Hamiltonian or stream function of the flow is given by,

\begin{equation}\label{eq:7}
\psi(x,y) = \frac{y^3}{3} - y - cos(x).
\end{equation}

It is a time-independent flow and the system is defined over the region $x\in(-\pi,\pi)$ and $y\in(-\pi,\pi)$.  The fluid particle trajectories are obtained by integrating the equations of motion.

\begin{equation}\label{eq:8}
\dot{x} = \frac{\partial\psi(x,y)}{dy}
\end{equation}

\begin{equation}\label{eq:9}
\dot{y} = -\frac{\partial\psi(x,y)}{dx}.
\end{equation}

\begin{figure}[htbp]
\centering
\includegraphics[width=8cm, height = 8cm]{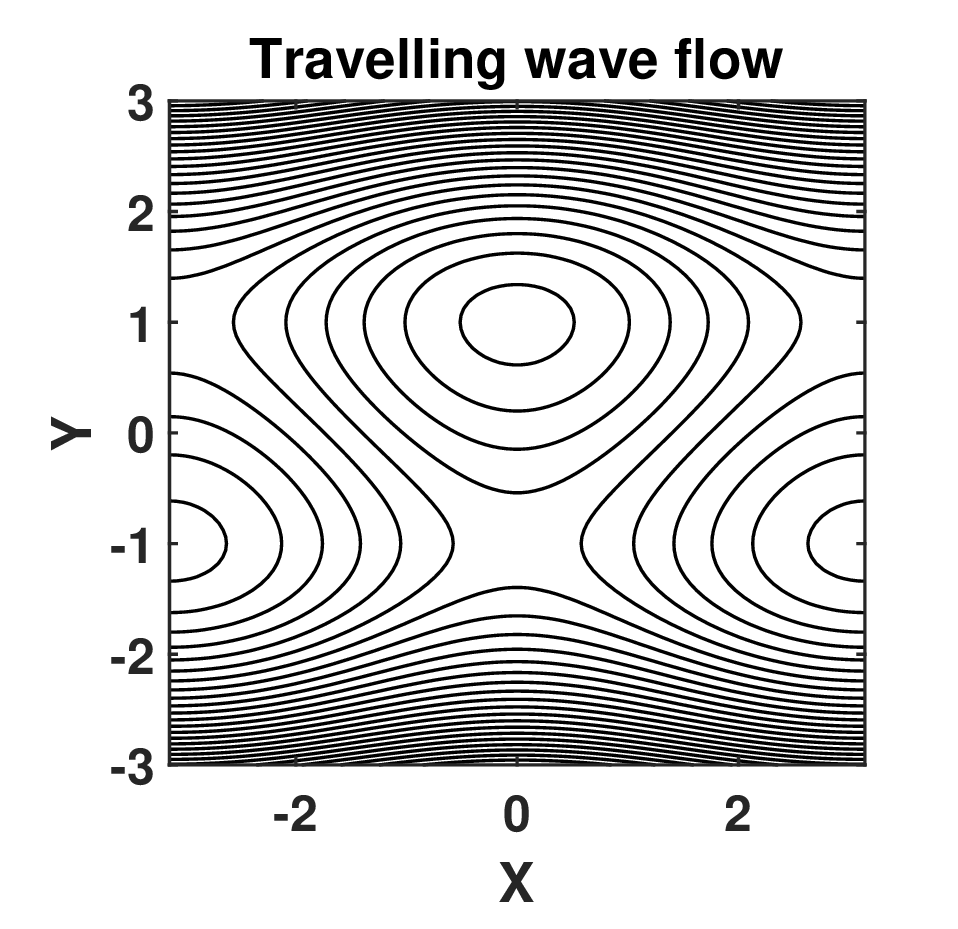} 
\caption{Streamlines are plotted for traveling wave flow in the domain $(-\pi,\pi)$ and $(-\pi, \pi)$.}     
\label{fig:1}
\end{figure}

In Fig. \ref{fig:1}, streamlines are plotted for the traveling wave flow which is an Eulerian field velocity using the above Eq. (\ref{eq:8}) and Eq. (\ref{eq:9}). The flow has two stable and two unstable fixed points.  The stable fixed points are (0,1) and ($\pi$,-1).  The unstable fixed points are (0,-1) and ($\pi$,1).  The fluid particles are trapped around the stable fixed points and these regions are separated by the free regions which are around the unstable fixed points. To get the flow field of the base flow we use initial conditions spread uniformly over the region $(-\pi,\pi)\times(-\pi,\pi)$. The resulting trajectory patterns reveal the nature of the flow and analyzing the resulting flow field gives a better understanding of the dynamics of the system, including the stability and instability of certain regions, and the potential for chaotic behavior near the separatrices.

\subsubsection*{\textbf{B. Fluid Tracers in a Traveling Wave Flow}}

Here, we studied the behavior of fluid tracer trajectories which are introduced in the fluid flow. As previously mentioned, initially the tracers are uniformly distributed within the defined spatial domain. The next step involves integrating these tracers over a specific time period to observe the flow patterns.  The specific integration time chosen plays a crucial role in shaping these patterns.

\begin{figure*}[htbp]
    \begin{subfigure}{0.32\linewidth}
        \centering
        \begin{overpic}[width=\linewidth, height =4.4 cm]{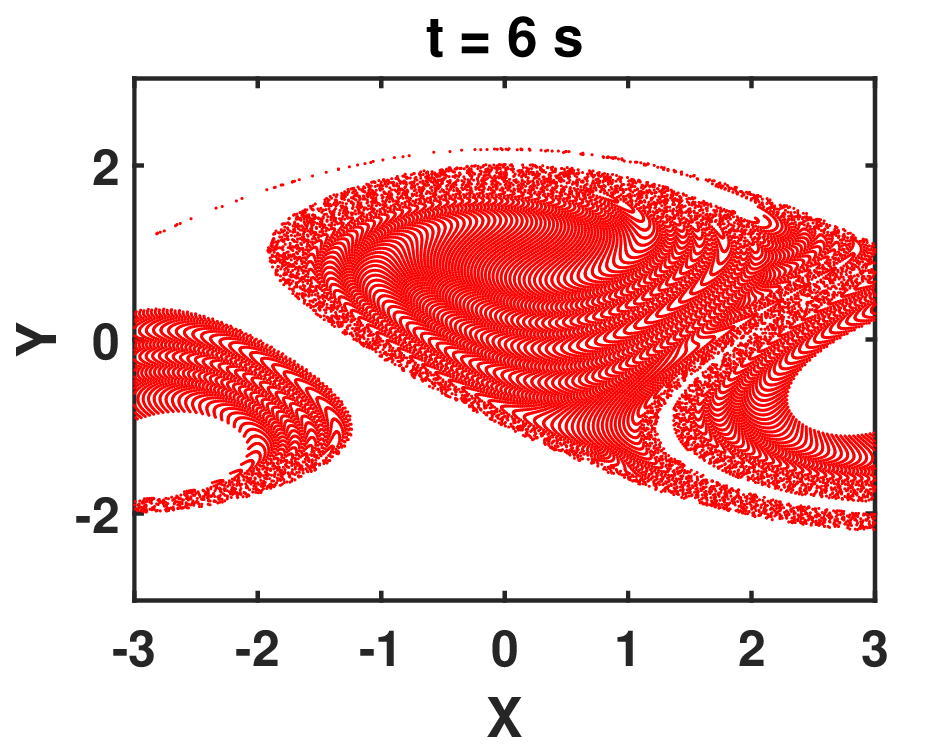} 
        \put(1,75){\textbf{(a)}}
        \end{overpic}     
        
         \label{fig:2a}
    \end{subfigure}%
     \begin{subfigure}{0.32\linewidth}
        \centering
        \begin{overpic}[width=\linewidth, height = 4.4cm]{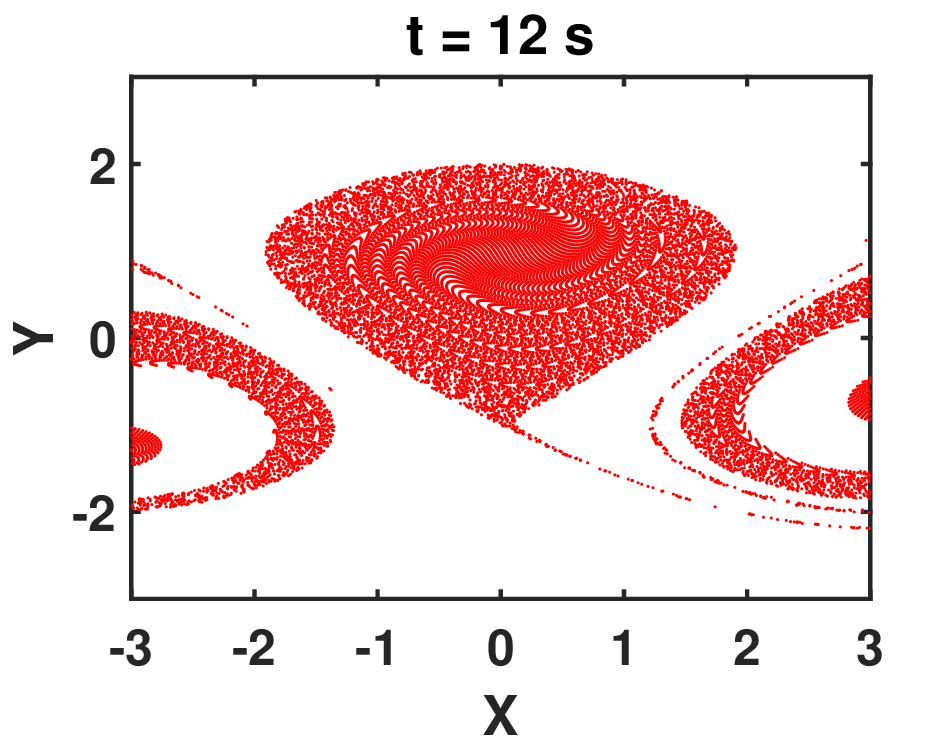}
        \put(1,75){\textbf{(b)}}
        \end{overpic}       
        
        \label{fig:2b}
    \end{subfigure}%
    \begin{subfigure}{0.32\linewidth}
        \centering
        \begin{overpic}[width=\linewidth, height =4.4cm]{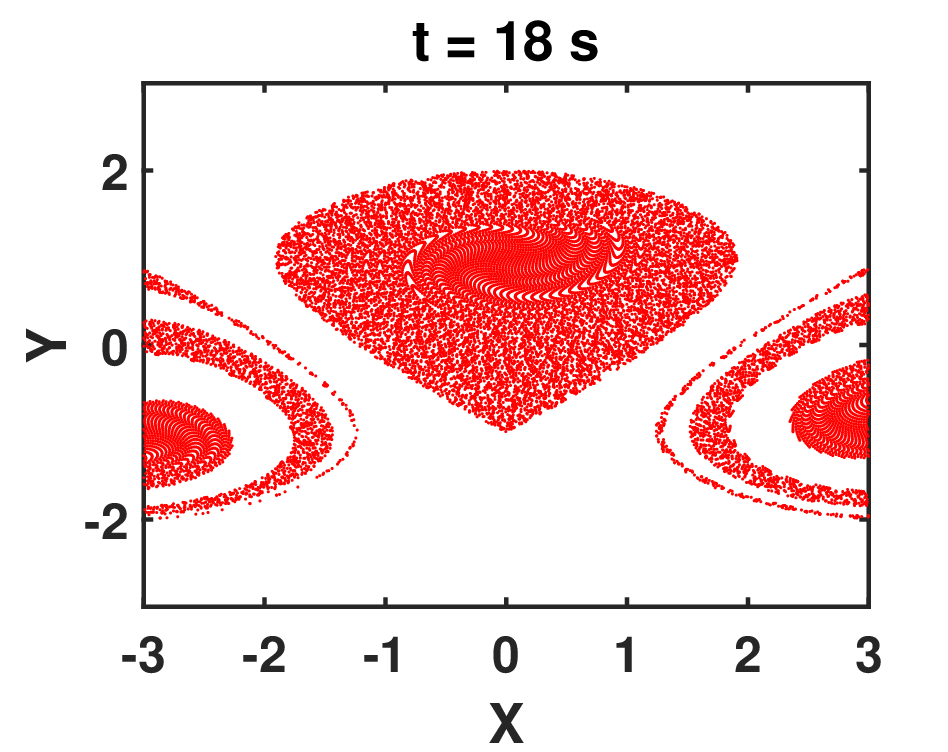}
        \put(1,75){\textbf{(c)}}
        \end{overpic}       
     
        \label{fig:2c}
    \end{subfigure}
    \caption{Lagrangian particles are initiated within the domain $x\in(-\pi,\pi)$ and $ y \in(-\pi,\pi)$ and are integrated over a time period.  Particle trajectories for different time integrations $t$ = 6 s, 12 s, and 18 s are plotted. The Lagrangian particles settle over time, and the patterns are clear at $t$ = 18 s.}
    \label{fig:2}
\end{figure*}

Fig. \ref{fig:2} illustrates the trajectories of fluid tracers, specifically non-inertial particles acting as fluid tracers for varying time integration intervals: $t$ = 6 s, 12 s, and 18 s which is shown in Fig. \ref{fig:2}(a), \ref{fig:2}(b) and \ref{fig:2}(c). Initially, the tracers are uniformly distributed across the domain. Upon integration, the fluid tracers undergo advection, leading to the gradual formation of distinct patterns, which become more pronounced by $t$ = 18 s.

\begin{figure*}
    \begin{subfigure}{0.32\linewidth}
        \centering
        \begin{overpic}[width=\linewidth, height = 4.5cm]{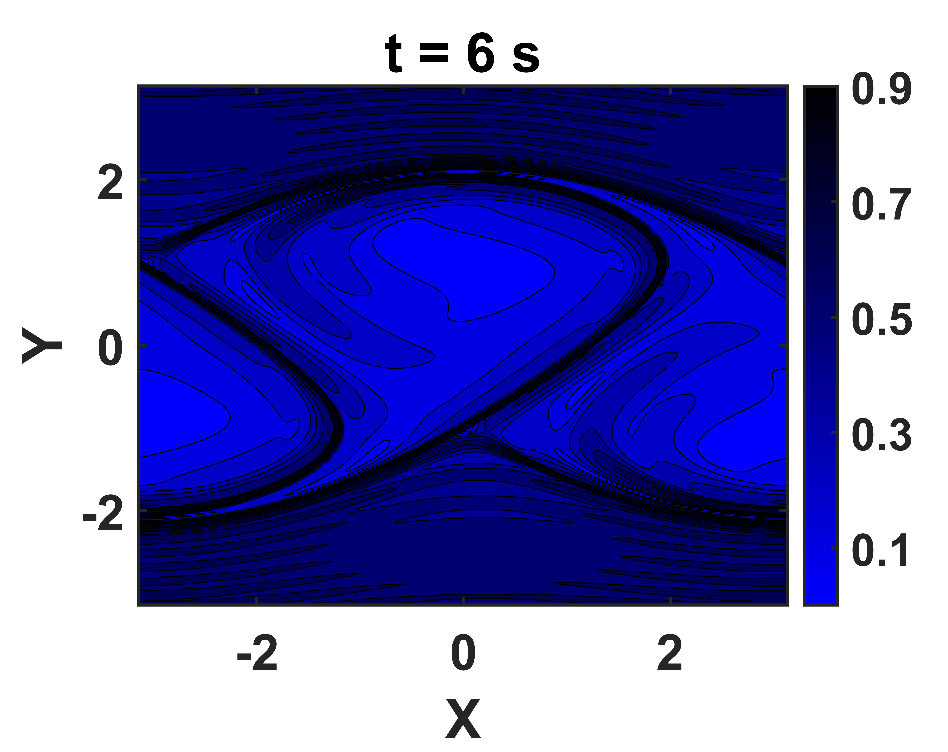} 
        \put(1,75){\textbf{(a)}}
        \end{overpic} 
        
        \label{fig:3a}   
        
    \end{subfigure}%
     \begin{subfigure}{0.32\linewidth}
        \centering
        \begin{overpic}[width=\linewidth, height = 4.5cm]{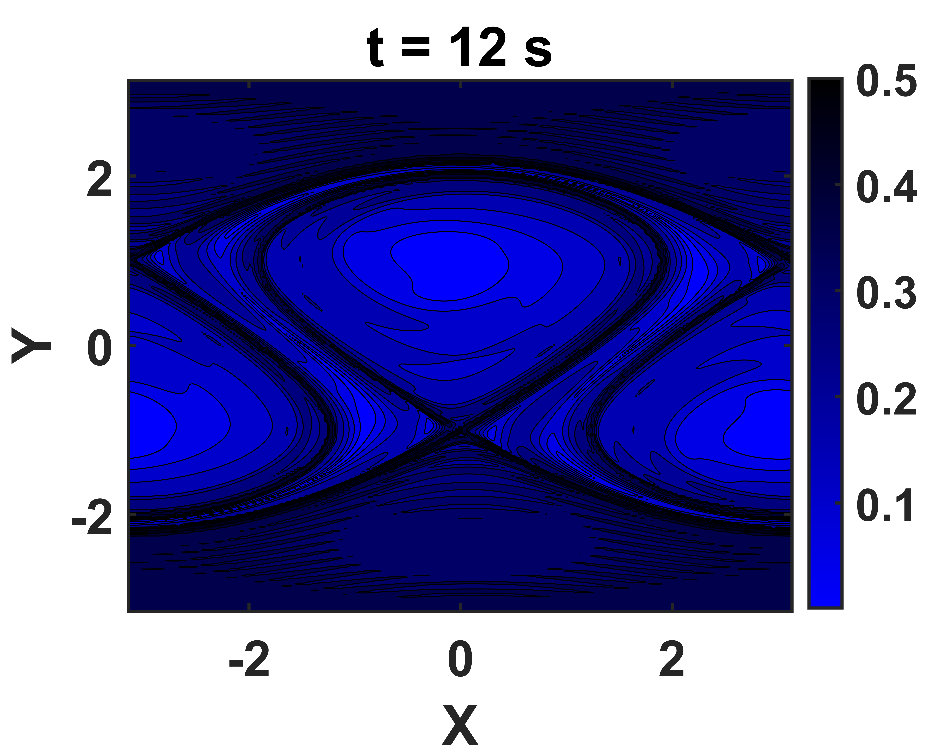} 
        \put(1,75){\textbf{(b)}}
        \end{overpic}   
  
        \label{fig:3b}  
        
    \end{subfigure}%
    \begin{subfigure}{0.32\linewidth}
        \centering
        \begin{overpic}[width=\linewidth, height = 4.5cm]{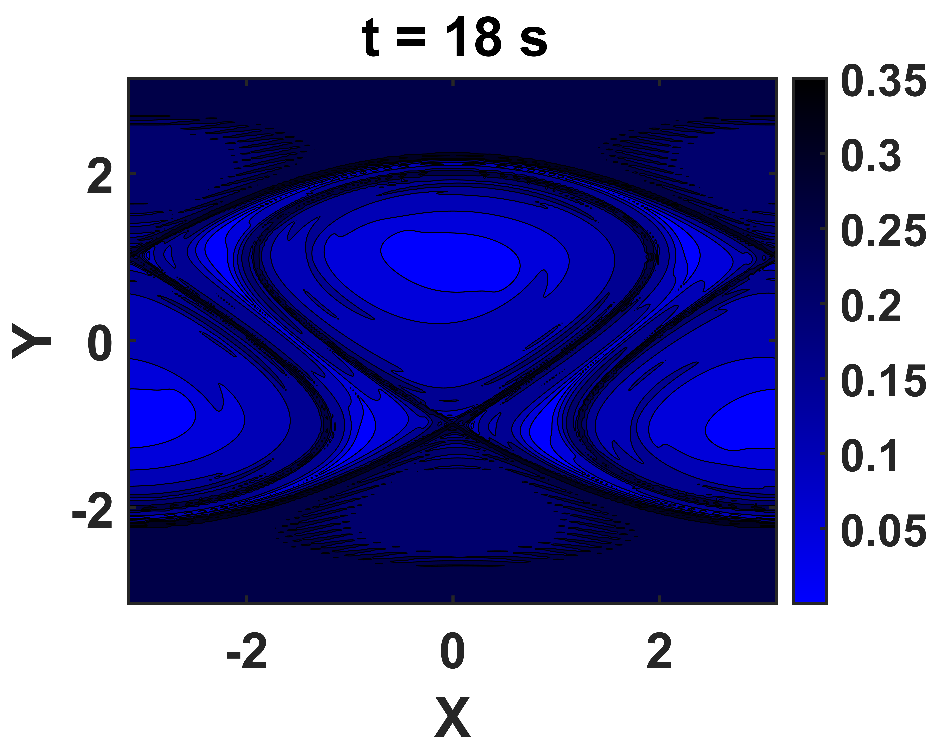} 
        \put(1,75){\textbf{(c)}}
        \end{overpic} 
    
        \label{fig:3c}  
       
    \end{subfigure}
    \caption{The positive finite time Lyapunov exponent (pFTLE) is plotted for the corresponding fluid trajectories(Fig:1).  Lagrangian particles are integrated forward in time and plotted for different time integrations $t$ = 6 s, 12 s, and 18 s. The dark black regions correspond to the largest eigenvalues, while the light blue regions correspond to the smallest eigenvalues. At $t$ = 8 s, the ridges are more prominent.}
    \label{fig:3}
\end{figure*}

Upon careful observation of these patterns, the primary focus lies in determining the largest Lyapunov exponent. This exponent is instrumental in identifying points within the flow where the stretching of neighboring trajectories is most pronounced. Consequently, we compute the finite-time Lyapunov exponent (FTLE) to quantify this behavior. The FTLE calculation involves evaluating the maximum singular value decomposition, which offers insights into regions of maximal stretching.

To pinpoint locations of fluid stretching, we not only analyze the FTLE values for forward integration in time but also engage in a reverse integration process, often referred to as negative FTLE analysis. This approach reveals areas where fluid tracers experience significant exponential stretching. In this context, the ridges of negative FTLE values correspond to regions characterized by such stretching phenomena within the fluid flow. Conversely, when integrating forward in time, the ridges of FTLE values guide us to attracting material lines, which contribute to our understanding of the flow's attractor dynamics. These analyses collectively enable a comprehensive insight into the intricate behavior of particles within the fluid flow, highlighting both regions of substantial stretching and the emergence of attracting material lines.

Fig. \ref{fig:3} illustrates the finite-time Lyapunov exponent (FTLE) field representing the behavior of fluid tracers over varying integration time intervals. The finite-time Lyapunov exponent (FTLE) is computed through numerical methods, and the visualization of fluid tracer behavior is achieved by advancing it forward in time, which we refer to as positive-time integration (pFTLE). These pFTLE ridges correspond to attracting material lines and also denote exponential stretching of the fluid flow when integrated in reverse time, termed negative-time integration (nFTLE). In Fig. \ref{fig:3}, it is apparent that as the integration time is extended, the density of FTLE ridges increases which can prominently be seen in Fig. \ref{fig:3}(a), \ref{fig:3}(b)  and \ref{fig:3}(c)  suggesting enhanced mixing among particles in proximity to these ridges.  A similar phenomenon is reported by Sudharsan et. al. \cite{sudharsan2016lagrangian}.

After identifying the ridges that represent the exponential stretching of the fluid flow, we did a computational analysis to validate our observations. We placed a group of particles which is a tracer, in two distinct regions: one associated with the largest eigenvalue (dark blue regions in Fig. \ref{fig:3}) and the other with the lowest eigenvalue (light blue regions in  Fig. \ref{fig:3}).

In the region with the largest eigenvalue (dark blue), neighboring particles exhibited exponential separation as the computational simulation progressed. Their trajectories diverged significantly from one another, indicating substantial stretching.

Conversely, when we positioned the group of particles in the region with the lowest eigenvalue (light blue region in Fig. \ref{fig:3}), particles nearby showed minimal separation throughout the simulation. They tended to remain near each other.

\begin{figure*}
    \begin{subfigure}{0.32\linewidth}
        \centering
        \begin{overpic}[width=\linewidth, height = 4.5cm]{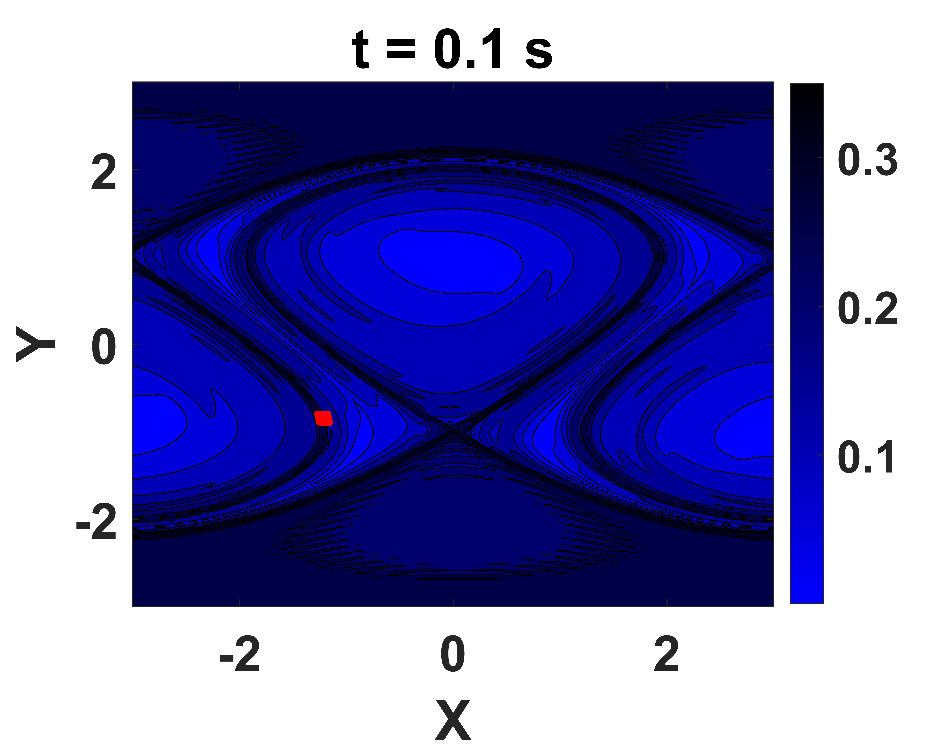}
        \put(1,75){\textbf{(a)}}
        \end{overpic} 
         
       \label{fig:4a}  
        
    \end{subfigure}%
     \begin{subfigure}{0.32\linewidth}
        \centering
        \begin{overpic}[width=\linewidth, height = 4.5cm]{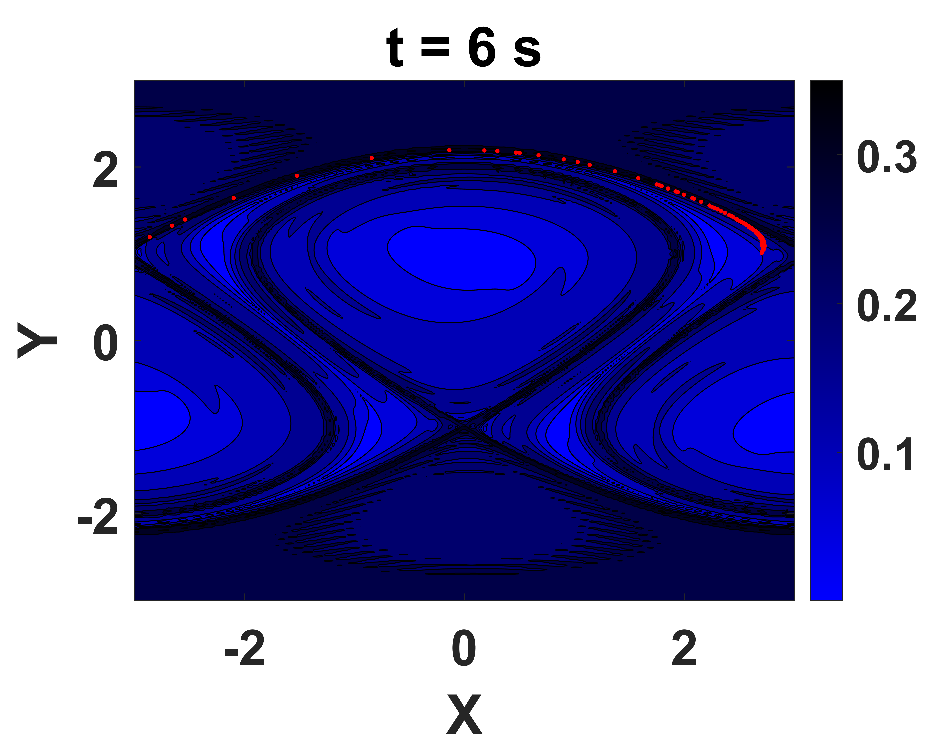}   
        \put(1,75){\textbf{(b)}}
        \end{overpic}   
        
       \label{fig:4b}
       
    \end{subfigure}%
    \begin{subfigure}{0.32\linewidth}
        \centering
         \begin{overpic}[width=\linewidth, height = 4.5cm]{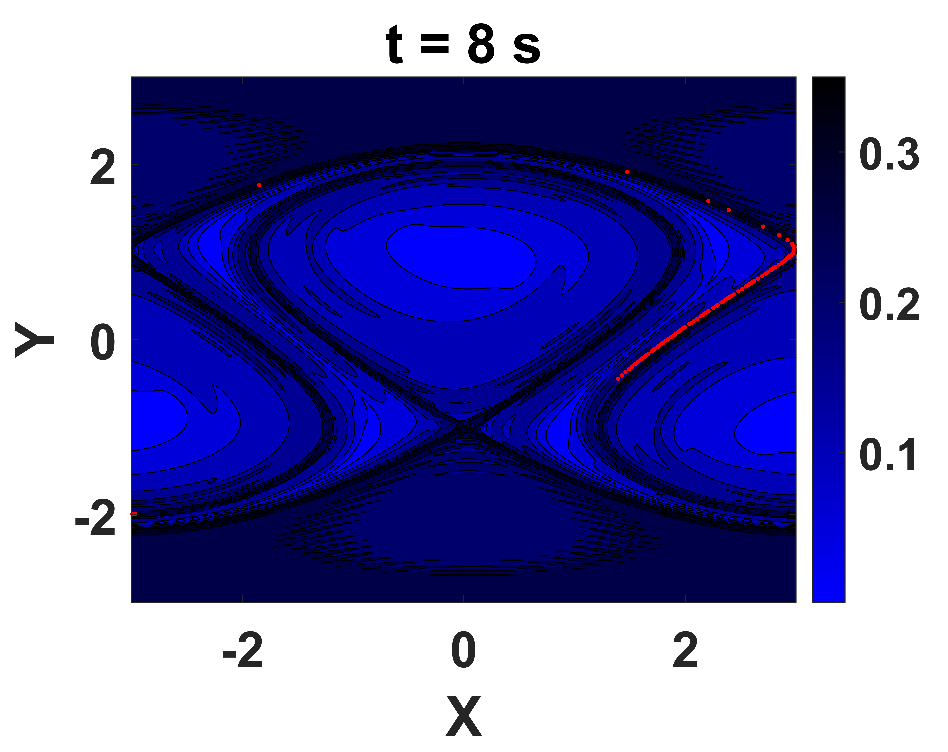}
        \put(1,75){\textbf{(c)}}
        \end{overpic}   
        
      \label{fig:4c}
       
    \end{subfigure}
    \caption{Red dots represent fluid particles initially placed in the dark black regions, which indicate the largest eigenvalues. These particles are positioned at coordinates $x$ = -1.242 to $x$ = -1.142 (interval 0.01) and $y$ = -0.992 to $y$ = -0.892 (interval 0.01). Over time, the particles stretch, move farther apart, and follow the path of the dark black regions, as shown in Fig. \ref{fig:4}(a), \ref{fig:4}(b), and \ref{fig:4}(c).}
    \label{fig:4}
\end{figure*}

\begin{figure*}
    \begin{subfigure}{0.32\linewidth}
        \centering
        \begin{overpic}[width=\linewidth, height = 4.5cm]{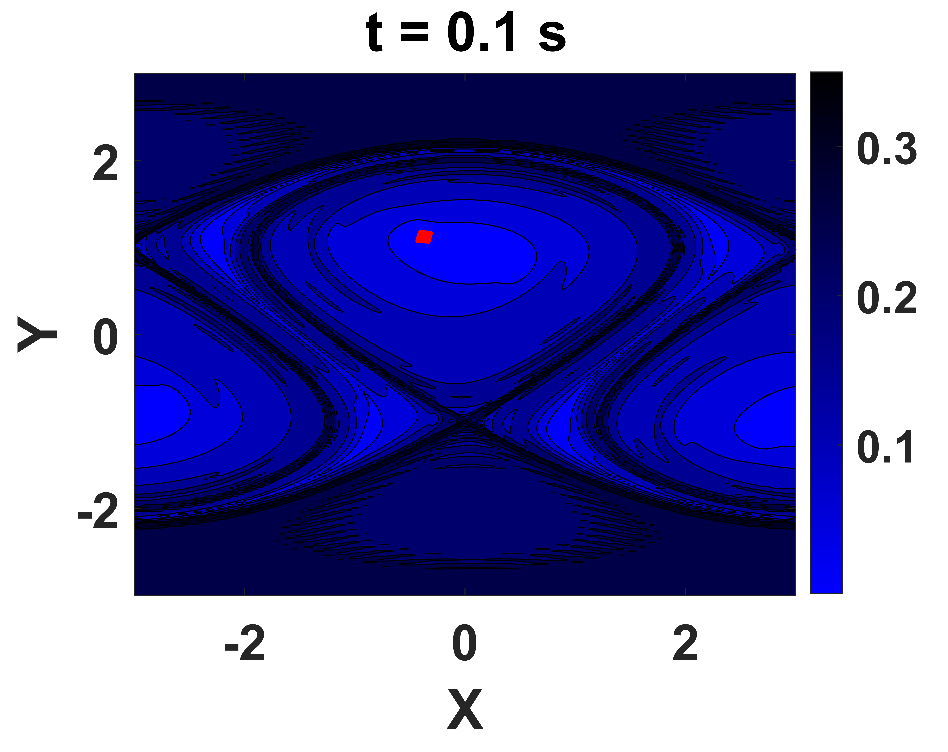}
        \put(1,75){\textbf{(a)}}
        \end{overpic}    
        
        \label{fig:5a}   
        
    \end{subfigure}%
     \begin{subfigure}{0.32\linewidth}
        \centering
         \begin{overpic}[width=\linewidth, height = 4.5cm]{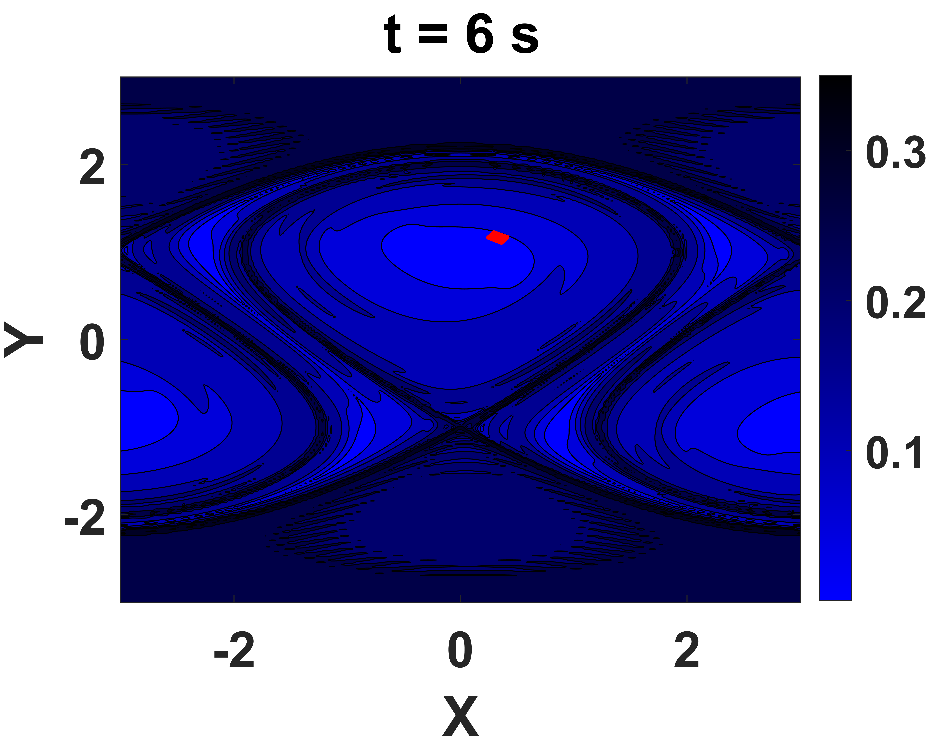}
        \put(1,75){\textbf{(b)}}
        \end{overpic}     
        
        \label{fig:5b}
       
    \end{subfigure}%
    \begin{subfigure}{0.32\linewidth}
        \centering
        \begin{overpic}[width=\linewidth, height = 4.5cm]{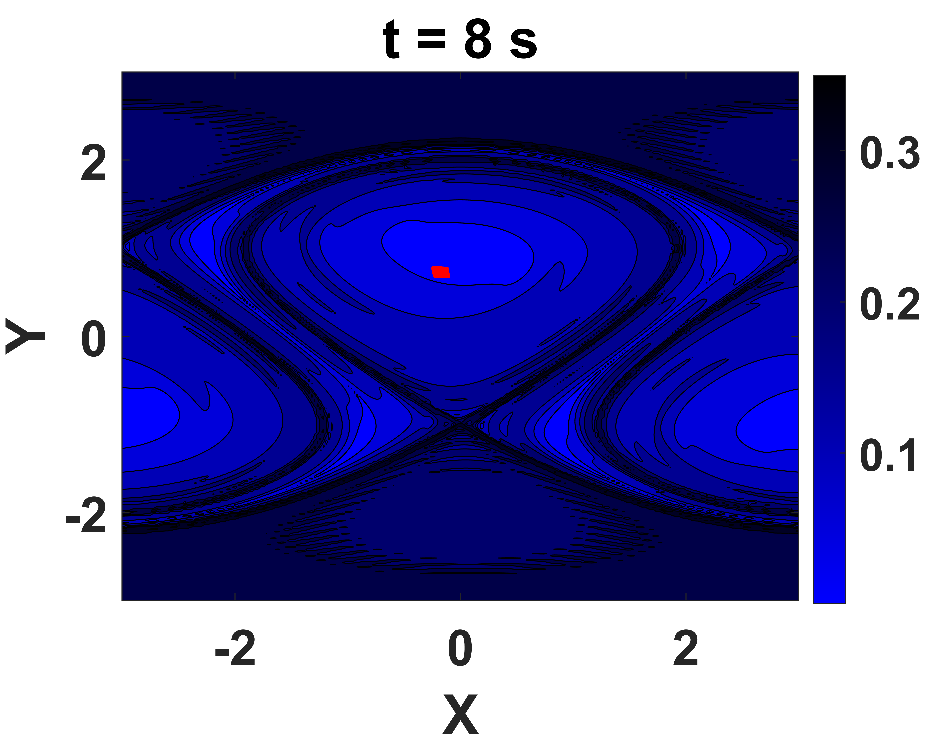}  
        \put(1,75){\textbf{(c)}}
        \end{overpic}   
        
        \label{fig:5c}
        
    \end{subfigure}
    \caption{Red dots represent fluid particles initially placed in the light blue regions, which indicate the smallest eigenvalues. These particles are positioned at coordinates $x$ = -0.442 to $x$ = -0.342 (interval 0.01) and $y$ = 1.033 to $y$ = 1.133 (interval 0.01). Over time, the particles remain close to each other with minimal separation, as shown in Fig. \ref{fig:5}(a), \ref{fig:5}(b), and \ref{fig:5}(c).}
    \label{fig:5}
\end{figure*}

In Fig. \ref{fig:4}, we initiated a cluster of particles at coordinates ( $x$ = -1.24159, $y$ = -0.991593) where it has the largest eigenvalue. Over a specified integration time, these nearby particles experienced exponential stretching, resulting in significant separation. Specifically, in Fig. \ref{fig:4}(a), the particles were initially placed at $t$ = 0.1 s at the coordinates mentioned and were tracked up to $t$ = 8 s. By the end of the 8th second, these particles had stretched significantly apart, a phenomenon clearly illustrated in the following Fig. \ref{fig:4}(b) and \ref{fig:4}(c).

In Fig. \ref{fig:5}, a cluster of particles was initially introduced at the coordinates ($x$ = -2.89159, $y$ = -0.866593). Throughout integration, it was observed that these nearby particles remained closely grouped, without experiencing exponential stretching between them and they moved together. Therefore, when a particle was introduced in the region where it has the lowest eigenvalue, the particles didn't exhibit significant stretching among themselves.  This phenomenon is distinctly portrayed in Fig. \ref{fig:5}(a) to \ref{fig:5}(c). In Fig. \ref{fig:5}(a), the particle positions at $t$ = 0.1 s are depicted, while Fig. \ref{fig:5}(b) shows their configuration at $t$ = 5 s. At $t$ = 8 s, it becomes evident that there was a notable absence of significant exponential stretching among the particles, as indicated in Fig. \ref{fig:5}(c).

In conclusion, we have observed the behavior of tracer particles upon their introduction into the fluid flow. We have analyzed to identify the regions of maximum stretching and regions that act as attracting material lines, which was achieved through the use of finite-time Lyapunov exponents.  Our next focus in the upcoming subsection will be to explore how inertial particles interact within the fluid flow when introduced into it.

\subsubsection*{\textbf{C. Inertial Particles in a Traveling Wave Flow}}

The behavior of inertial particles is investigated and the preferential concentrations are studied. Now, we are initiating a study of the behavior of inertial particles, specifically bubbles, and aerosols, under different combinations of $R$ and $St$ values.

\begin{figure*}
    \begin{subfigure}{0.32\linewidth}
        \centering
        \begin{overpic}[width=\linewidth, height = 4.5cm]{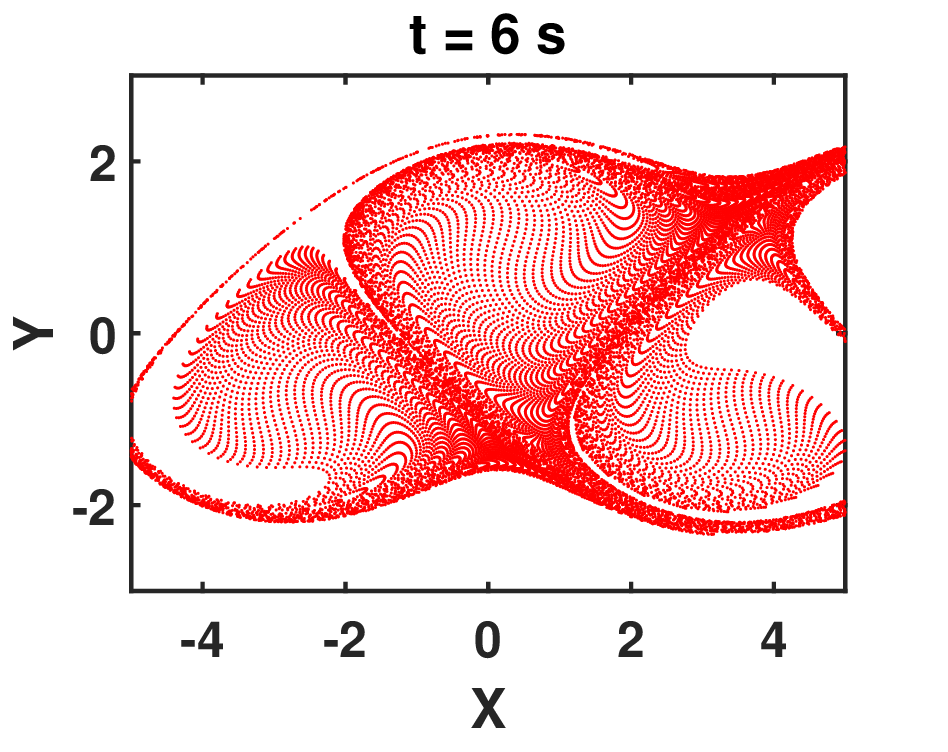} 
        \put(1,75){\textbf{(a)}}
        \end{overpic} 
       \label{fig:6a}   
        
    \end{subfigure}%
     \begin{subfigure}{0.32\linewidth}
        \centering
        \begin{overpic}[width=\linewidth, height = 4.5cm]{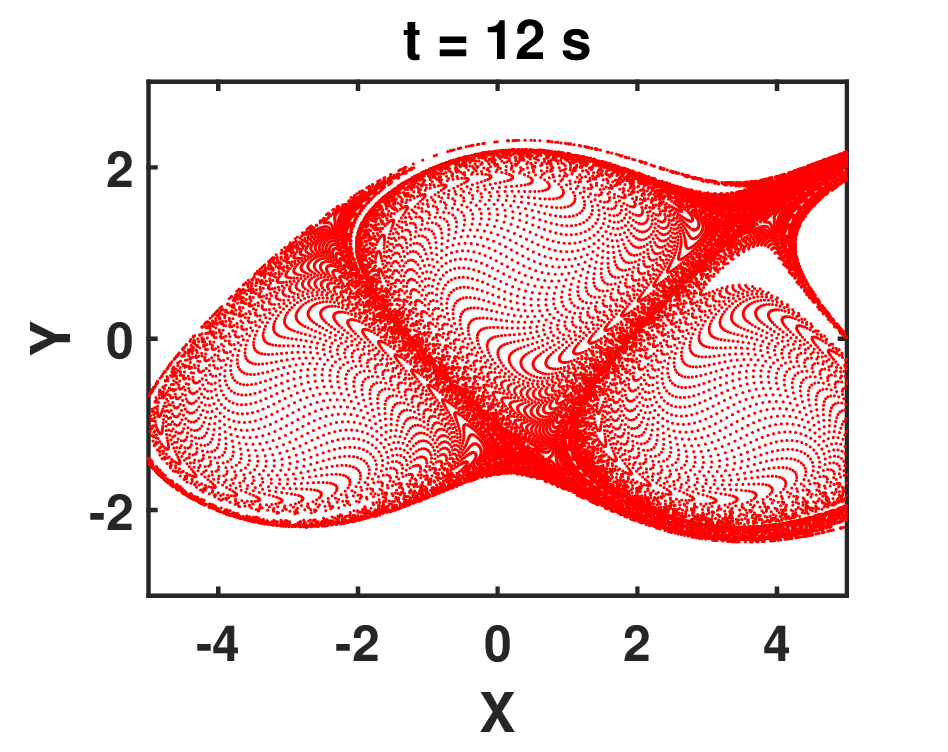}
        \put(1,75){\textbf{(b)}}
        \end{overpic} 
        \label{fig:6b}
       
    \end{subfigure}%
    \begin{subfigure}{0.32\linewidth}
        \centering
        \begin{overpic}[width=\linewidth, height = 4.5cm]{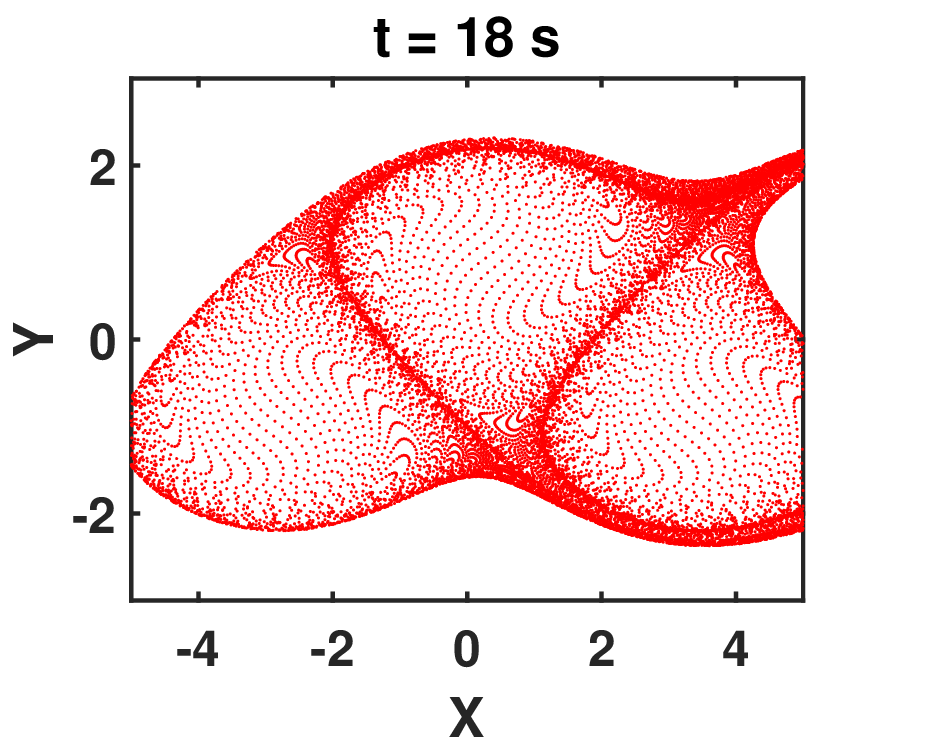} 
        \put(1,75){\textbf{(c)}}
        \end{overpic}       
        \label{fig:6c}
        
    \end{subfigure}
    \caption{Inertial particles (aerosols) are introduced in the fluid flow and their trajectories are plotted for various time integrations, specifically $t$ = 6 s, 12 s, and 18 s.  The other parameters are $R$ = 0 and $St$ = 0.1.  As time progresses, the particles are segregated along the attracting manifolds}
    \label{fig:6}
\end{figure*}

The primary goal is to uncover the Lagrangian coherent structures associated with inertial particles, and one of the commonly used methods to gain insights into fluid mechanics is through flow visualization \cite{babiano2000dynamics}. Within the framework of finite-time Lyapunov exponents (FTLE), two distinct features emerge: attracting FTLEs, which are derived from negative integration time, and repelling FTLEs, obtained through positive integration time. 

By utilizing Eq. (\ref{eq:3}), one can advect the inertial particles by integrating both their position and velocity. Fig. \ref{fig:6} provides a visual representation of the paths taken by aerosol particles (with $R$ = 0) at different time points while maintaining constant values for $R$ and $St$, specifically $St$ = 0.1 and $R$ = 0. As time progresses, these aerosol particles attract towards the attracting manifolds within the fluid which is evident in Fig. \ref{fig:6}(a) - \ref{fig:6}(c).

\begin{figure*}
    \begin{subfigure}{0.32\linewidth}
        \centering
        \begin{overpic}[width=\linewidth, height = 4.5cm]{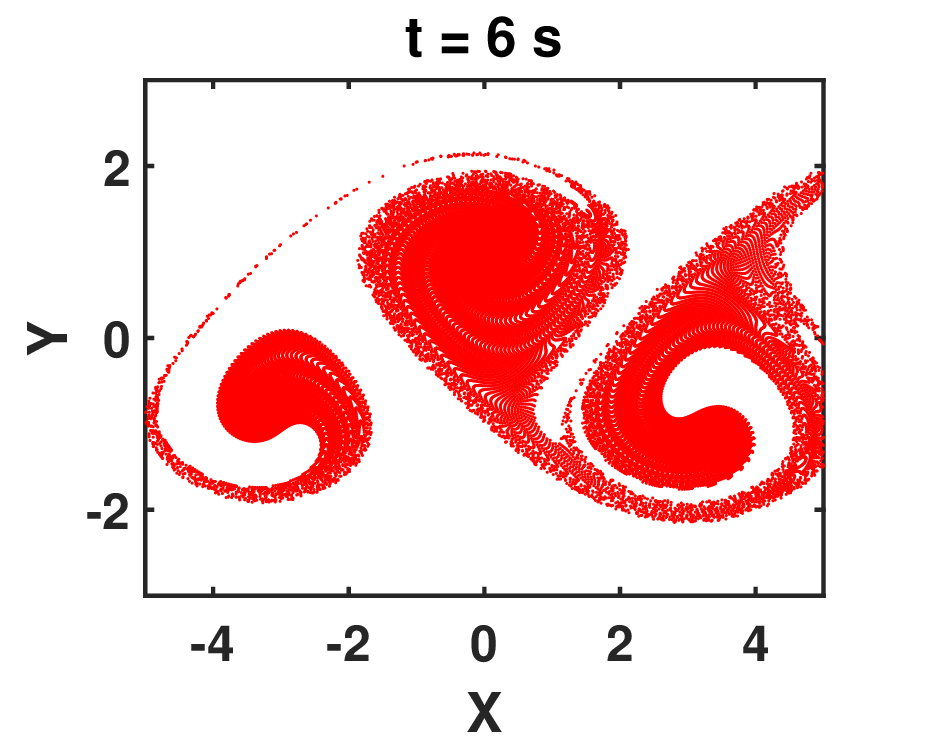}  
        \put(1,75){\textbf{(a)}}
        \end{overpic}  
       
        \label{fig:7a}
        
    \end{subfigure}%
     \begin{subfigure}{0.32\linewidth}
        \centering
        \begin{overpic}[width=\linewidth, height = 4.5cm]{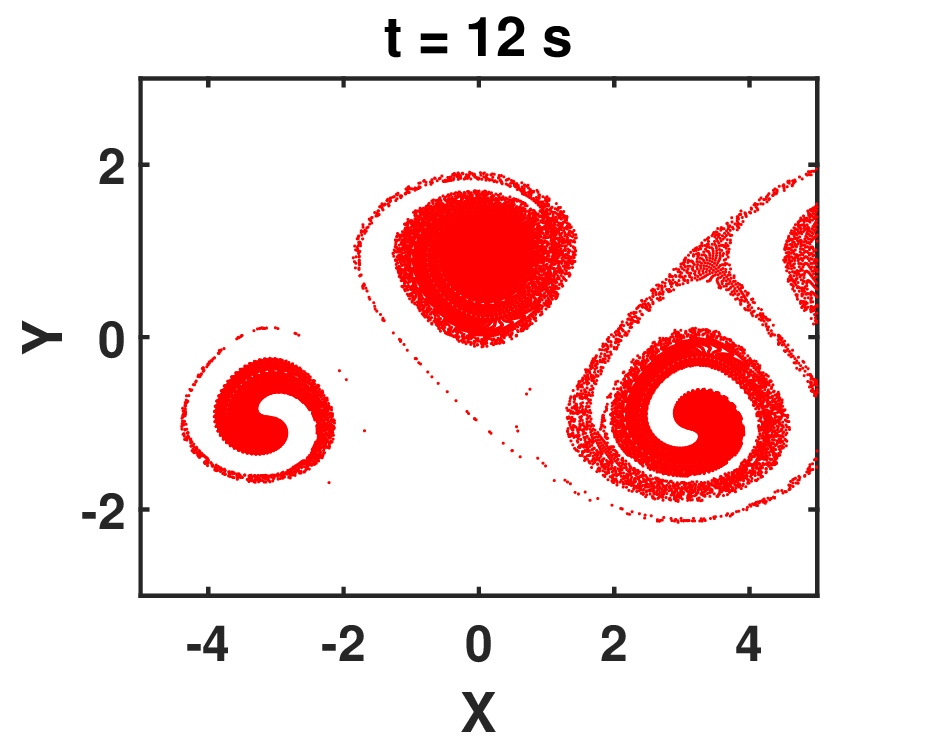} 
        \put(1,75){\textbf{(a)}}
        \end{overpic} 
           
       \label{fig:7b}
       
    \end{subfigure}%
    \begin{subfigure}{0.32\linewidth}
        \centering
        \begin{overpic}[width=\linewidth, height = 4.5cm]{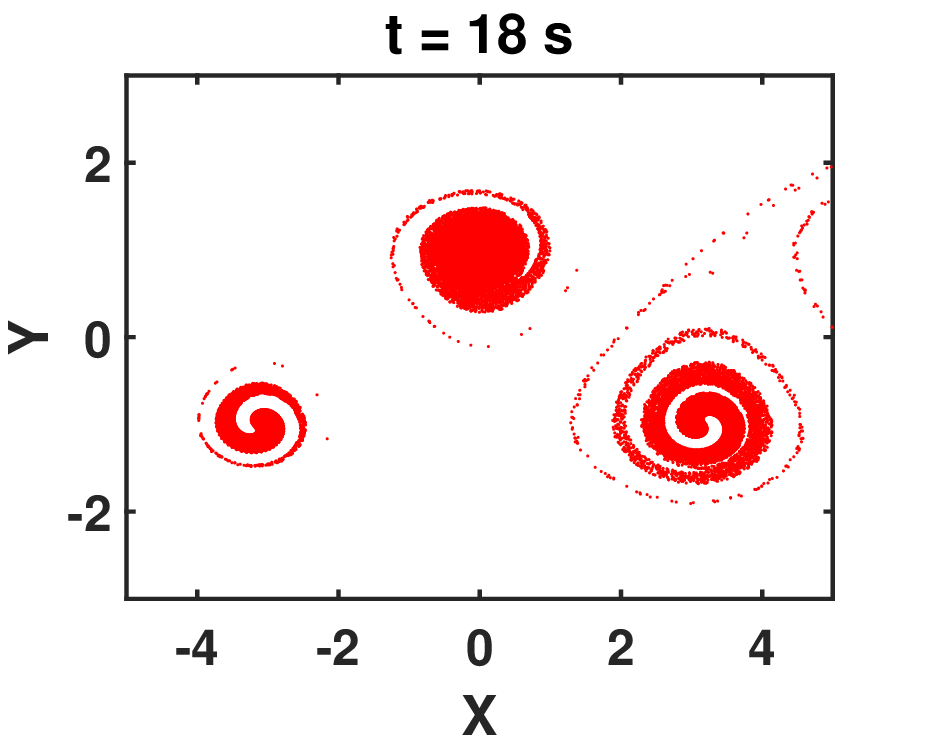}   
        \put(1,75){\textbf{(c)}}
        \end{overpic} 
          
       \label{fig:7c}
        
    \end{subfigure}
    \caption{Inertial particles (bubbles) are introduced in the fluid flow and their trajectories are plotted for various time integrations, specifically $t$ = 6 s, 12 s, and 18 s.  The other parameters are $R$ = 0 and $St$ = 0.1.  As time progresses, the particles are repelled from the attracting manifolds}
    \label{fig:7}
\end{figure*}

Similarly, Fig. \ref{fig:7} shows the visual representation of the trajectories followed by bubble particles (with $R$ = 1) at different integration time intervals by keeping the values of $St$ and $R$ constant, specifically $R$ = 1 and $St$ = 0.1. Over time, these bubble-like particles are pushed away or repelled from the attracting manifolds which is evident in Fig. \ref{fig:7}(a) - \ref{fig:7}(c). This phenomenon can be elucidated by attributing it to preferential concentration effects.  As we discussed earlier the material lines of nFTLE accurately pinpoint and reveal the structures.  These results imply that the utilization of weakly aerosol particles with $R$ values slightly below 2/3 may enhance the visualization of attracting flow structures. While neutrally buoyant particles are typically preferred in flow visualization studies to minimize interference with the flow, our findings suggest that slightly denser particles with $R$ values just below 2/3 exhibit a stronger inclination toward the underlying attracting FTLE, compared to neutral particles.

\begin{figure*}
    \begin{subfigure}{0.32\linewidth}
        \centering
        \begin{overpic}[width=\linewidth, height = 4.5cm]{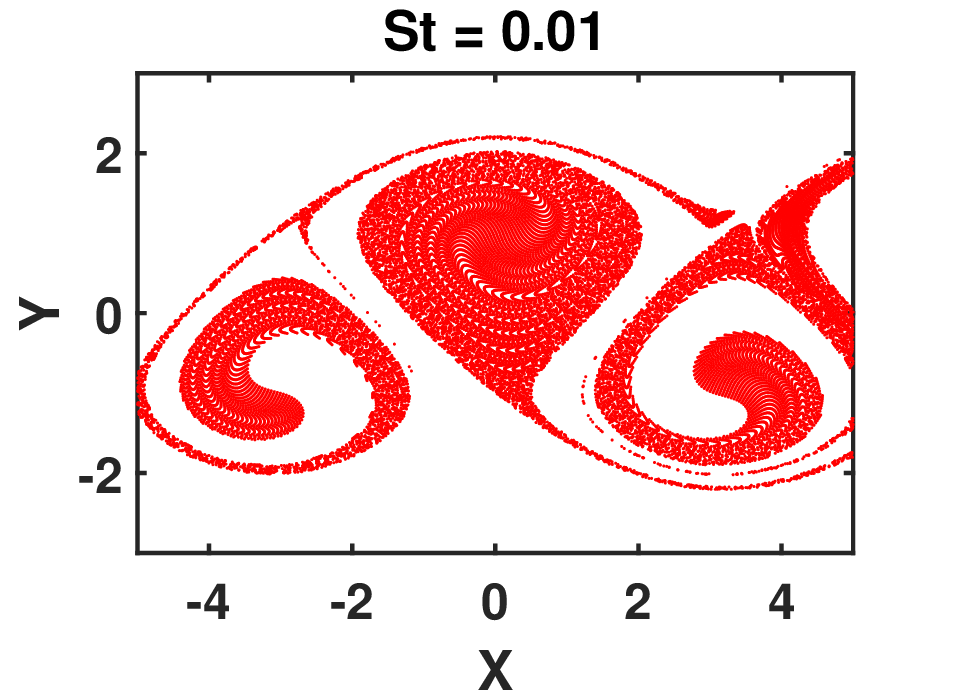}
        \put(1,75){\textbf{(a)}}
        \end{overpic}
          
       \label{fig:8a}   
        
    \end{subfigure}%
     \begin{subfigure}{0.32\linewidth}
        \centering
        \begin{overpic}[width=\linewidth, height = 4.5cm]{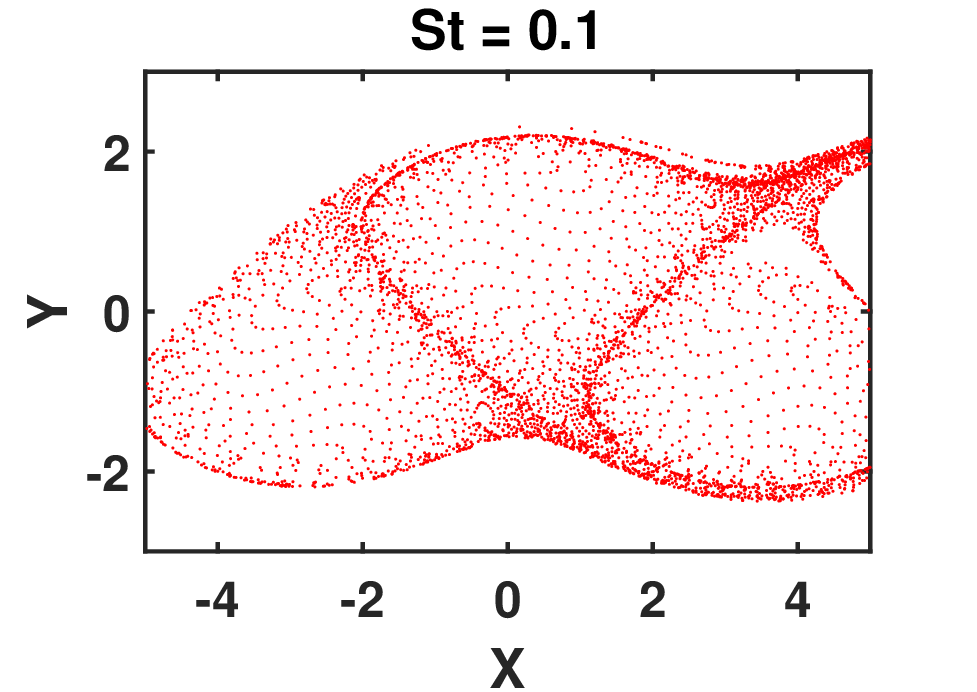}
        \put(1,75){\textbf{(b)}}
        \end{overpic}
             
       \label{fig:8b}
       
    \end{subfigure}%
    \begin{subfigure}{0.32\linewidth}
        \centering
        \begin{overpic}[width=\linewidth, height = 4.5cm]{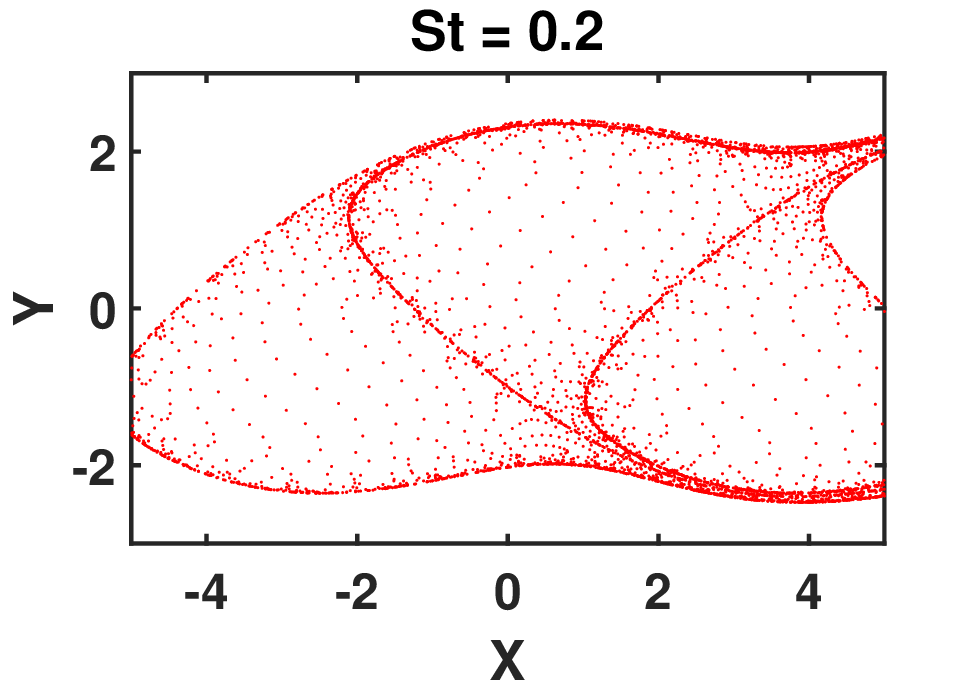}
        \put(1,75){\textbf{(c)}}
        \end{overpic}
             
       \label{fig:8c}
        
    \end{subfigure}
    \caption{Aerosol particle trajectories (with $R$ = 0)  are plotted for Stokes number $St$ = 0.01, 0.1 and 0.2 at $t$ = 12 s. As the Stokes number increases, the particles are attracted to the ridges. The figures illustrate the dissipation of phase space with higher Stokes number}
    \label{fig:8}
\end{figure*}

\begin{figure*}
    \begin{subfigure}{0.32\linewidth}
        \centering
        \begin{overpic}[width=\linewidth, height = 4.5cm]{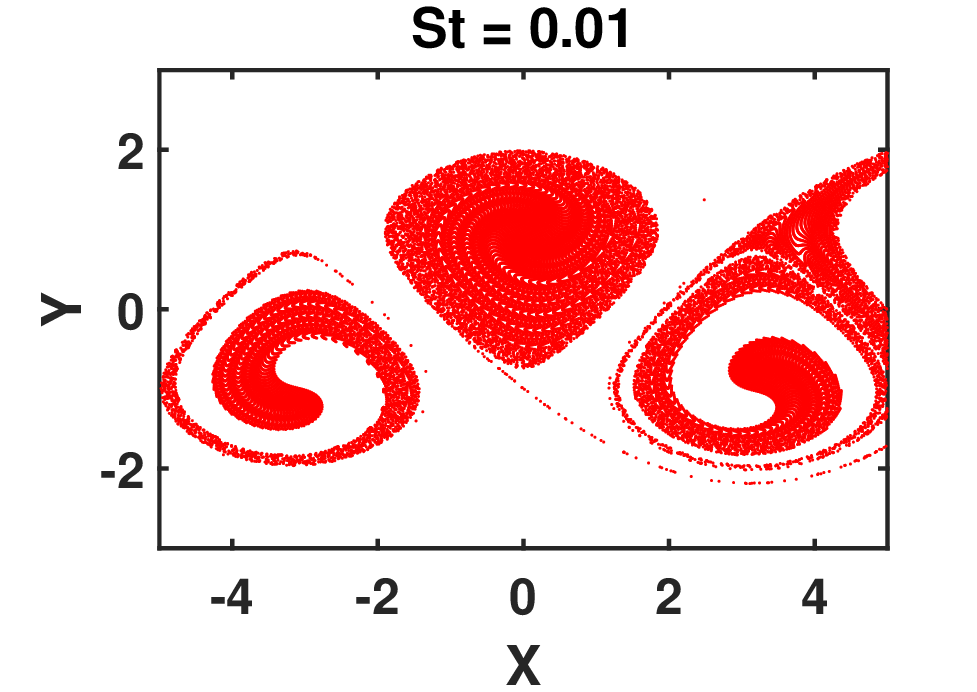}
        \put(1,75){\textbf{(a)}}
        \end{overpic}
              
       \label{fig:9a}
        
    \end{subfigure}%
     \begin{subfigure}{0.32\linewidth}
        \centering
        \begin{overpic}[width=\linewidth, height = 4.5cm]{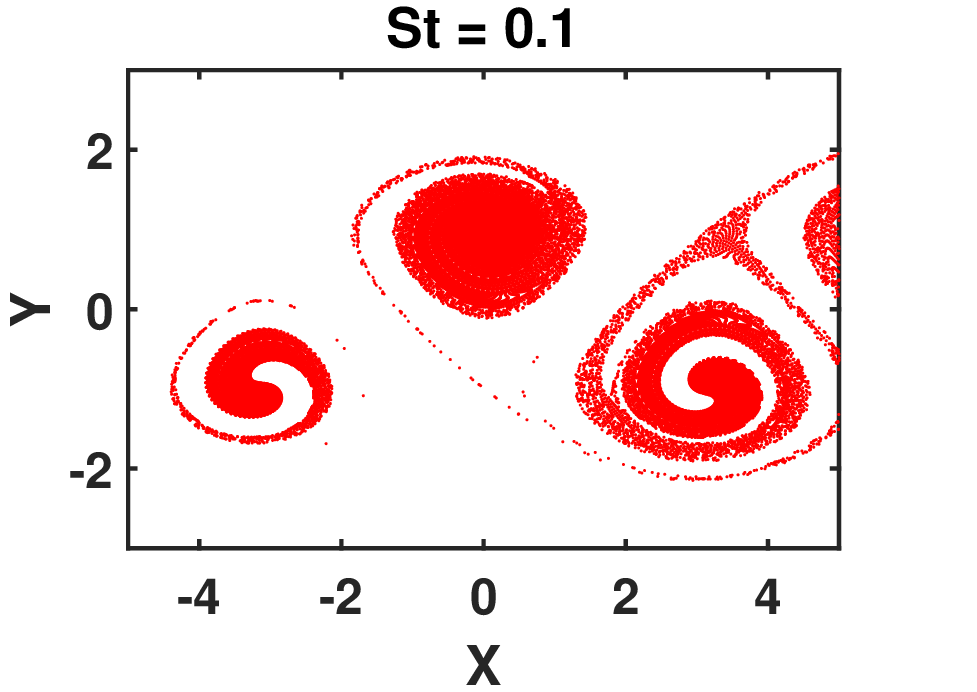} 
        \put(1,75){\textbf{(b)}}
        \end{overpic}   
         
       \label{fig:9b}
       
    \end{subfigure}%
    \begin{subfigure}{0.32\linewidth}
        \centering
        \begin{overpic}[width=\linewidth, height = 4.5cm]{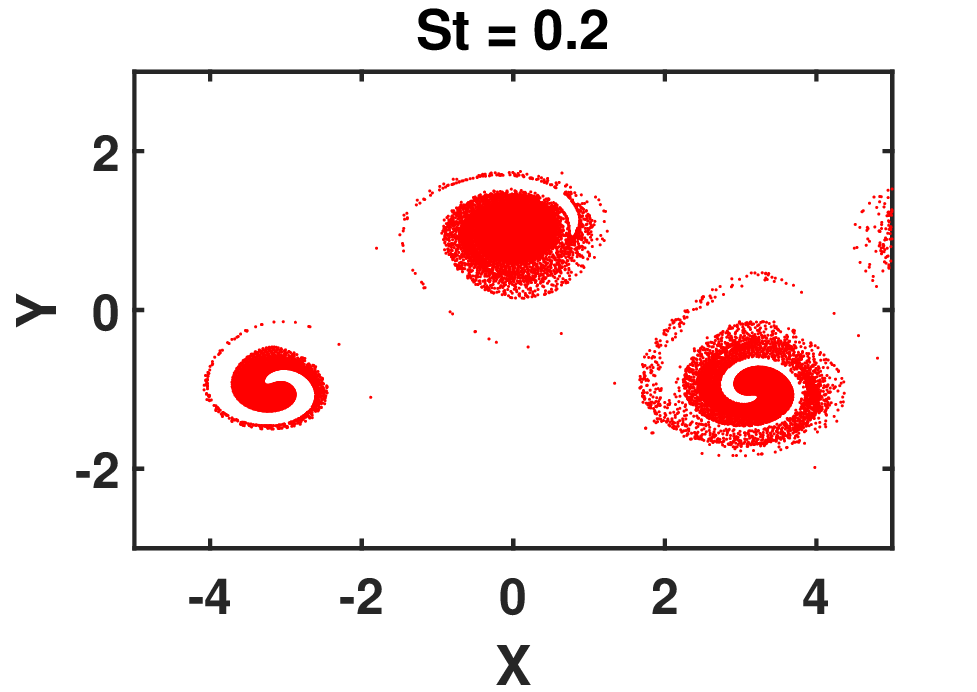}
        \put(1,75){\textbf{(c)}}
        \end{overpic}  
          
       \label{fig:9c}
        
    \end{subfigure}
    \caption{Bubble particle trajectories (with $R$ = 1)  are plotted for Stokes number $St$ = 0.01, 0.1 and 0.2 at $t$ = 12 s. As the Stokes number increases, the particles are pushed away from the ridges. The figures illustrate the more pronounced dispersal of phase space.}
    \label{fig:9}
\end{figure*}

In Fig. \ref{fig:8}(a), \ref{fig:8}(b), and \ref{fig:8}(c), we present the behavior of aerosol particle positions at time $t$ = 12 s while varying the Stokes number ($St$) from 0.01 to 0.2. Notably, when $St$ is at its lowest value of 0.01, the behavior of these particles resembles that of tracer particles, moving along with the flow. However, as $St$ increases to 0.2, a substantial dissipation of phase space becomes evident, characterized by significant divergence.

\begin{figure*}
    \begin{subfigure}{0.32\linewidth}
        \centering
        \begin{overpic}[width=\linewidth, height = 4.5cm]{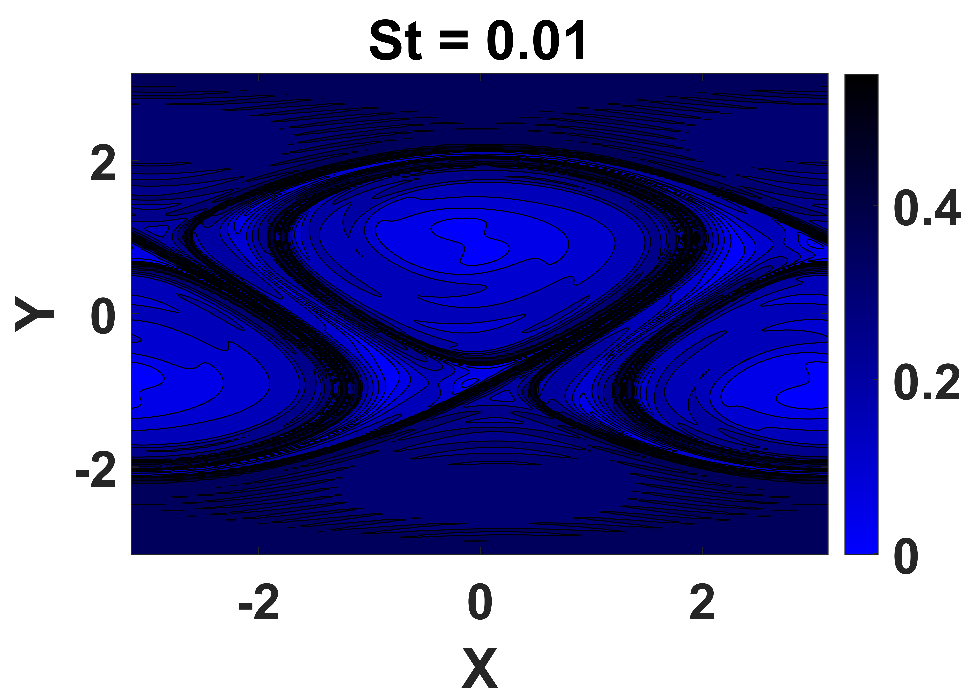} 
        \put(1,75){\textbf{(a)}}
        \end{overpic} 
           
       \label{fig:10a}
        
    \end{subfigure}%
     \begin{subfigure}{0.32\linewidth}
        \centering
        \begin{overpic}[width=\linewidth, height = 4.5cm]{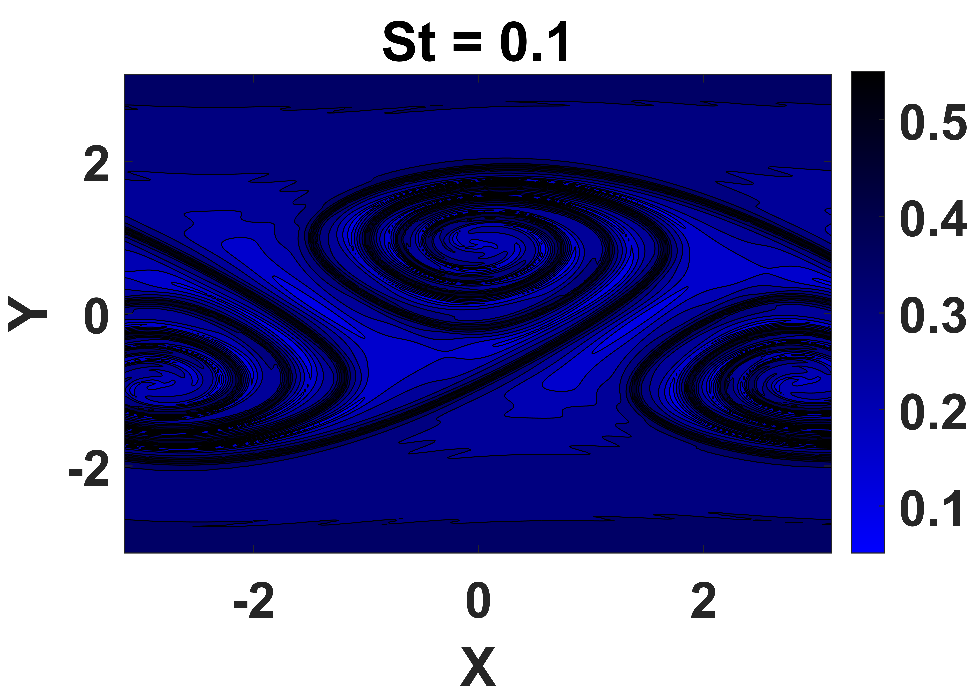}  
        \put(1,75){\textbf{(b)}}
        \end{overpic} 
            
       \label{fig:10b}
       
    \end{subfigure}%
    \begin{subfigure}{0.32\linewidth}
        \centering
        \begin{overpic}[width=\linewidth, height = 4.5cm]{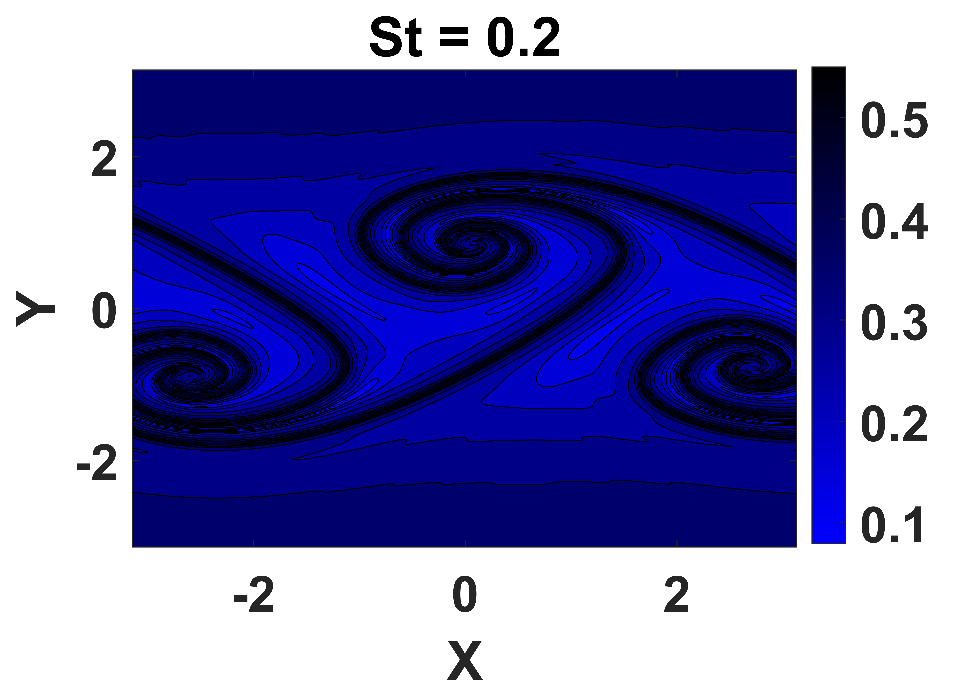}
        \put(1,75){\textbf{(c)}}
        \end{overpic} 
                 
       \label{fig:10c}
        
    \end{subfigure}
    \caption{Positive iFTLE fields are plotted for aerosol particles (with $R$ = 0) for varying Stokes numbers ($St$ = 0.01, 0.1, and 0.2) at $t$ = 12 s. The inertial particles (aerosols) are integrated forward in time, with ridges in the contour plots indicating material lines undergoing exponential stretching. As time increases, the number of ridges multiply and become more pronounced.}
    \label{fig:10}
\end{figure*}

The corresponding inertial finite-time Lyapunov exponents (iFTLEs) are plotted in Fig. \ref{fig:10}(a), \ref{fig:10}(b), and \ref{fig:10}(c) for aerosol particles. In this context, the inertial particles are integrated forward in time, and the ridges in the contour plots signify material lines experiencing exponential stretching.

Figures  \ref{fig:9}(a),  \ref{fig:9}(b), and  \ref{fig:9}(c) display the behavior of bubble particles' positions at $t$ = 12 s while varying $St$ from 0.01 to 0.2. These particles, in contrast to aerosols, are repelled from the ridges, which is evident in Fig. \ref{fig:9}. Moreover, as the value of $St$ is increased, bubble particles are completely pushed away from the ridges, showcasing a more pronounced dispersal in phase space.

\begin{figure*}
    \begin{subfigure}{0.32\linewidth}
        \centering
        \begin{overpic}[width=\linewidth, height = 4.5cm]{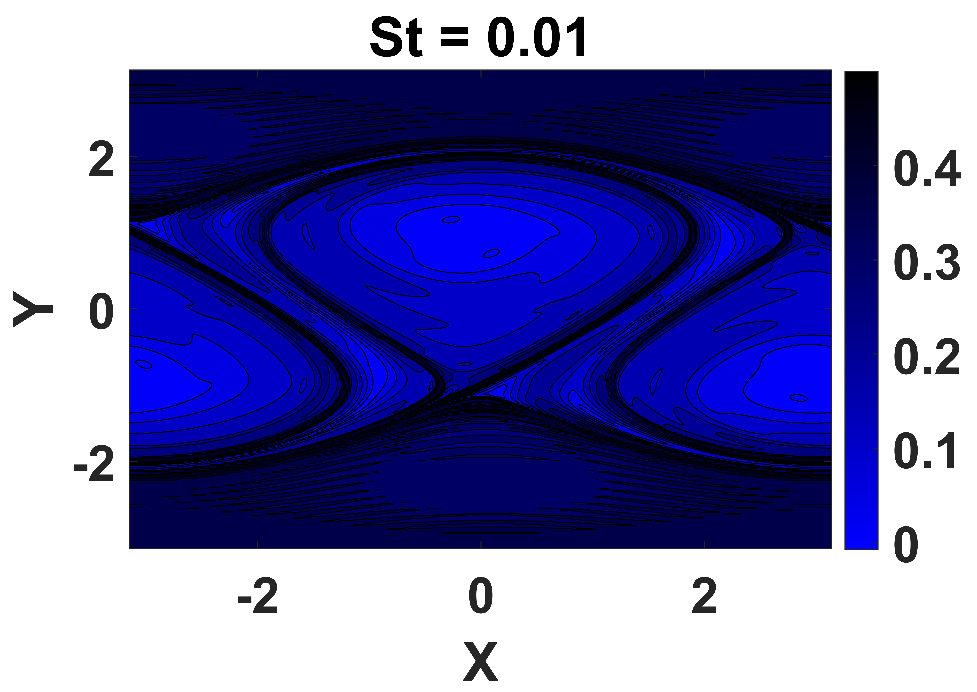}
        \put(1,75){\textbf{(a)}}
        \end{overpic}
              
        \label{fig:11a}
        
    \end{subfigure}%
     \begin{subfigure}{0.32\linewidth}
        \centering
        \begin{overpic}[width=\linewidth, height = 4.5cm]{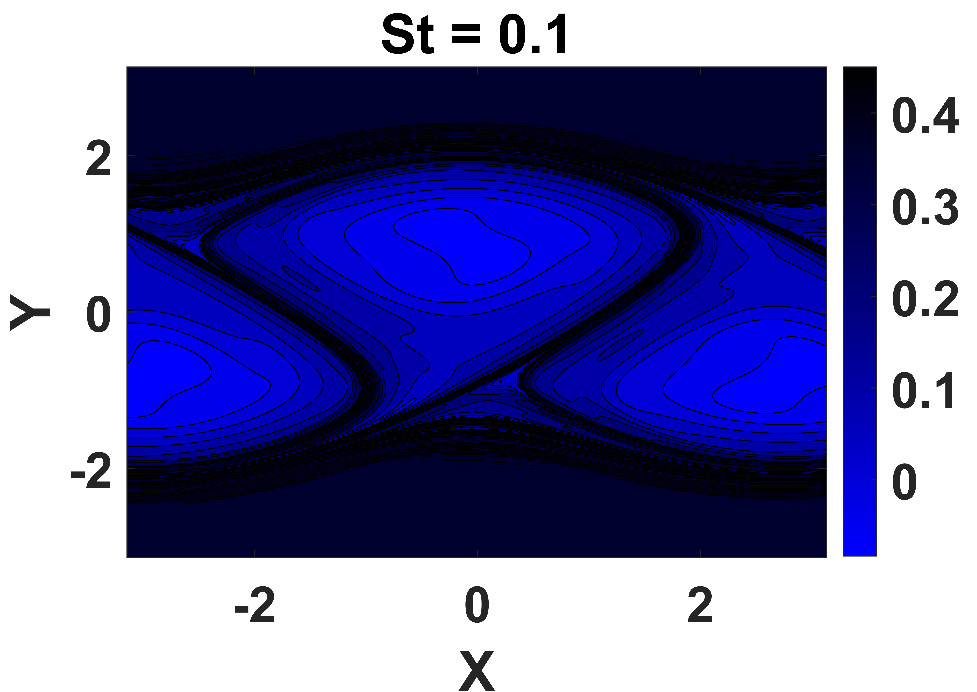} 
        \put(1,75){\textbf{(b)}}
        \end{overpic}
              
         \label{fig:11b}
       
    \end{subfigure}%
    \begin{subfigure}{0.32\linewidth}
        \centering
        \begin{overpic}[width=\linewidth, height = 4.5cm]{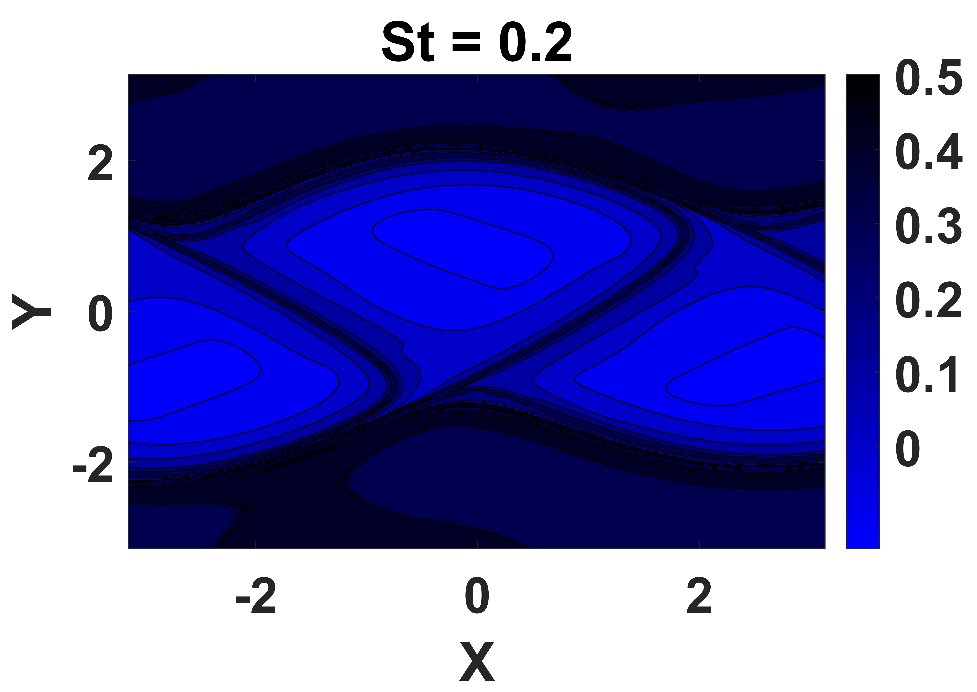} 
        \put(1,75){\textbf{(c)}}
        \end{overpic}
             
        \label{fig:11c}
        
    \end{subfigure}
    \caption{Positive iFTLE fields are plotted for bubble particles (with $R$ = 1) for varying Stokes numbers ($St$ = 0.01, 0.1 and 0.2) at $t$ = 12 s. The inertial particles (bubbles) are integrated forward in time, with ridges in the contour plots indicating material lines undergoing exponential stretching. As time increases the ridges are relatively fewer.}
    \label{fig:11}
\end{figure*}

These observations allow for the segregation of inertial particles based on the Stokes number ($St$). Specifically, for higher values of $St$, the particles are extracted from the flow region near the attractor. Additionally, when the Stokes number is increased for aerosol particles, the iFTLE ridges multiply, resulting in more pronounced ridges (as shown in Fig. \ref{fig:10}(a) to \ref{fig:10}(c)). In contrast, bubble particles exhibit relatively fewer iFTLE ridges (as observed in Fig. \ref{fig:11}(a) to \ref{fig:11}(c)). Therefore, iFTLE values serve as indicators of the exponential stretching of material lines and can be interpreted as a measure of particle mixing.

Overall, it is evident that increasing the Stokes number leads to improved mixing for aerosols ($R$ = 0), while the opposite is true for bubbles ($R$ = 1) particles. This outcome underscores the fact that optimal mixing occurs at different Stokes numbers for bubbles and aerosols.

In addition to the iFTLE fields, we have generated surface plots, which vividly display the regions with the highest peaks corresponding to the largest eigenvalues and the regions with the lowest peaks indicative of the smallest eigenvalues. The primary purpose of these surface plots is to provide further support for the findings from the iFTLE analysis.

\begin{figure*}
    \begin{subfigure}{0.32\linewidth}
        \centering
         \begin{overpic}[width=\linewidth, height = 4.5cm]{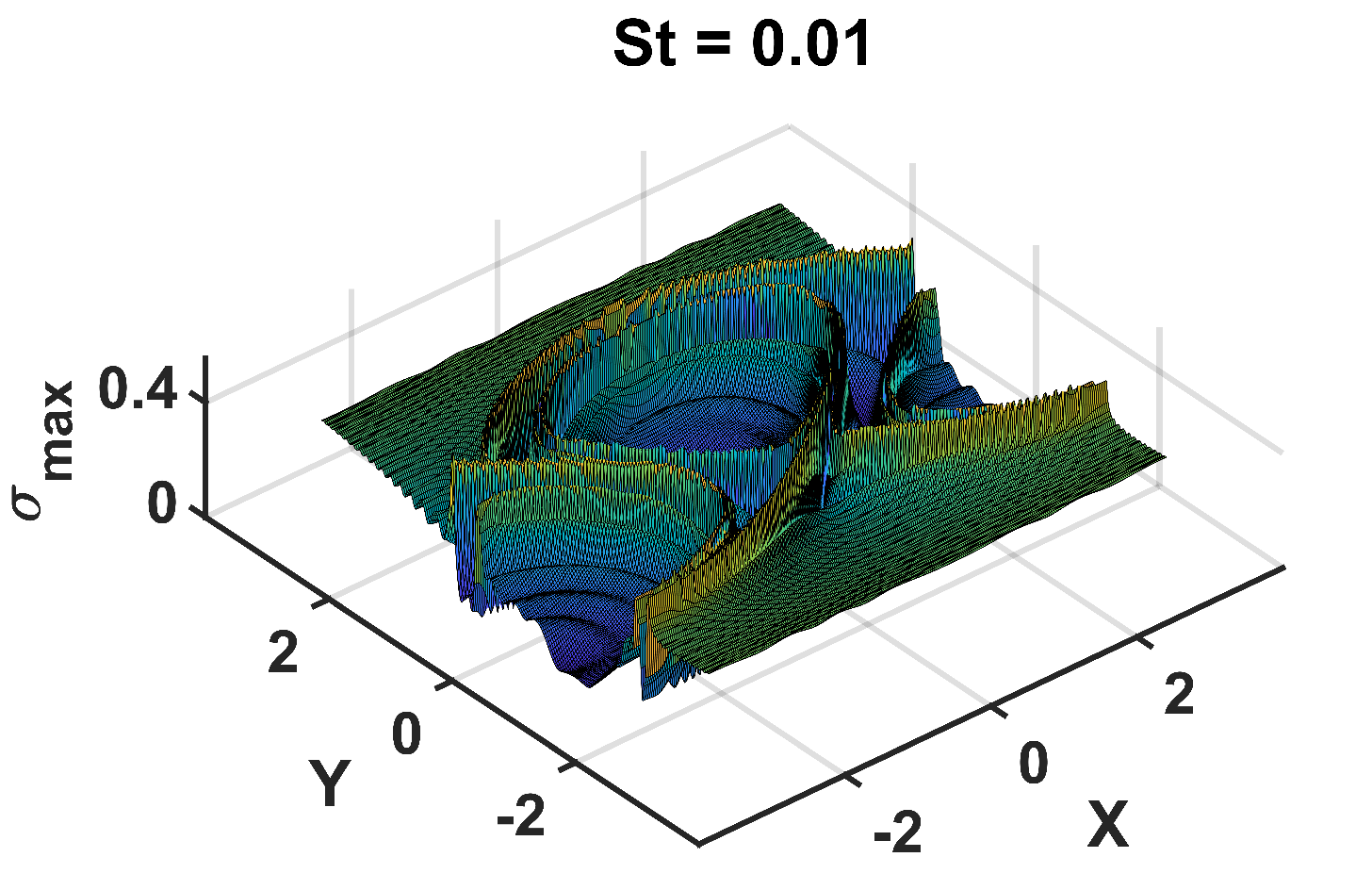} 
        \put(1,75){\textbf{(a)}}
        \end{overpic}
            
        \label{fig:12a}
        
    \end{subfigure}%
     \begin{subfigure}{0.32\linewidth}
        \centering
        \begin{overpic}[width=\linewidth, height = 4.5cm]{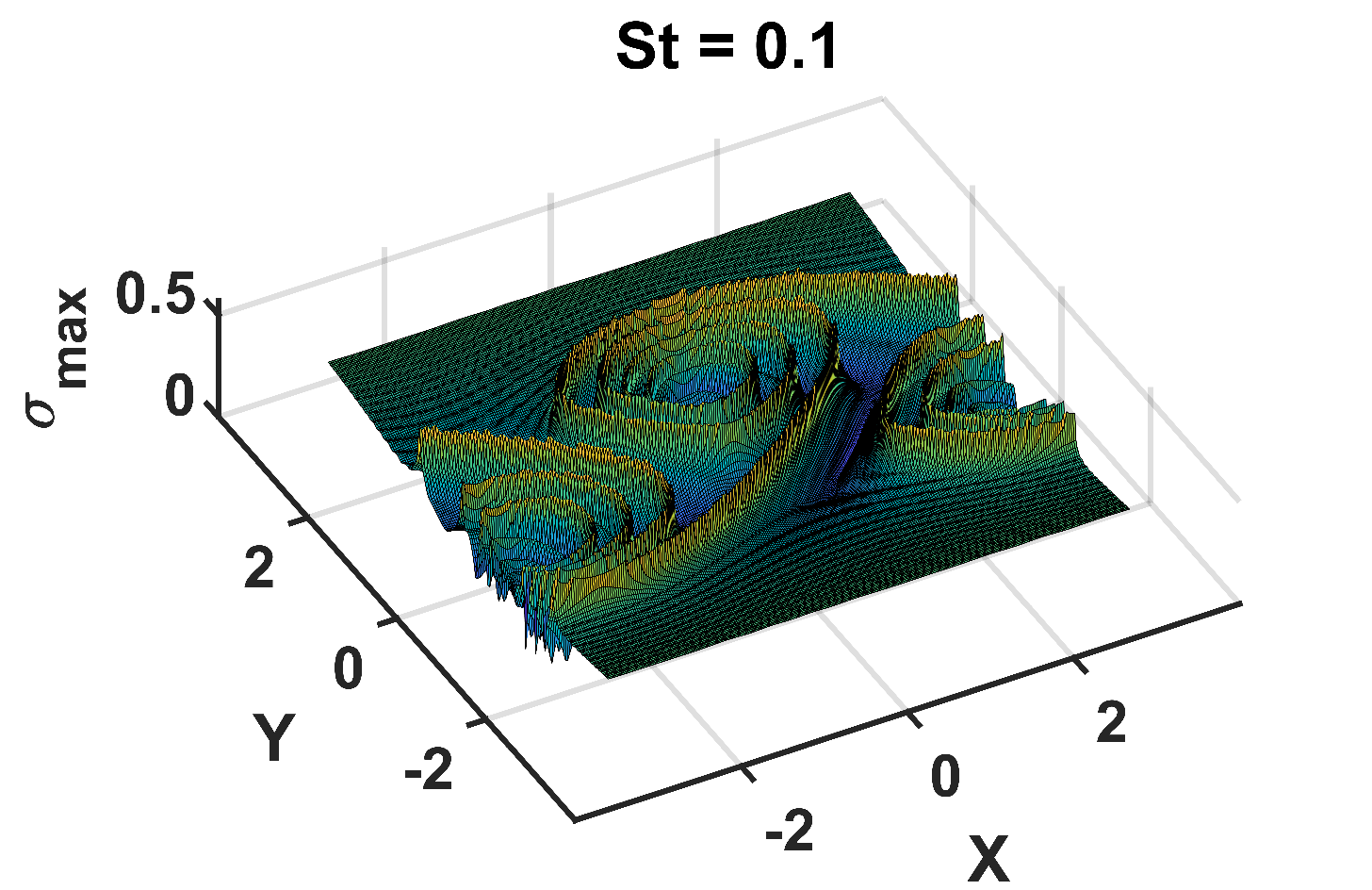}  
        \put(1,75){\textbf{(b)}}
        \end{overpic}   
         
        \label{fig:12b}
       
    \end{subfigure}%
    \begin{subfigure}{0.32\linewidth}
        \centering
        \begin{overpic}[width=\linewidth, height = 4.5cm]{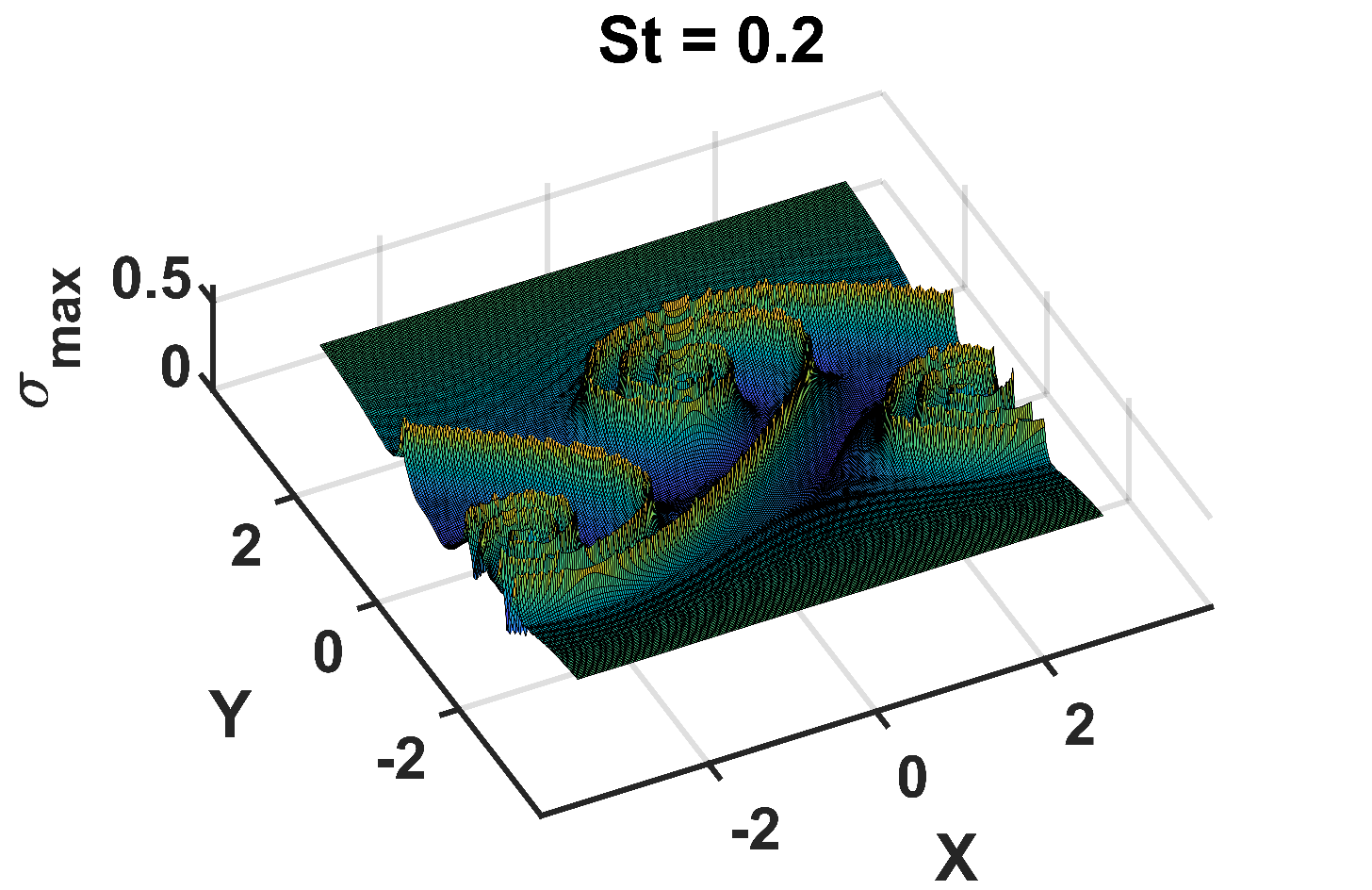}   
        \put(1,75){\textbf{(c)}}
        \end{overpic}  
           
        \label{fig:12c}
        
    \end{subfigure}
    \caption{Surface plots for the aerosol particles ($R$ = 0) for varying Stokes number say $St$ = 0.01, 0.1, and 0.2 at t =12 s are plotted. The highest peaks indicate the largest eigenvalues and the shortest peaks indicate the smallest eigenvalues.}
    \label{fig:12}
\end{figure*}

\begin{figure*}
    \begin{subfigure}{0.32\linewidth}
        \centering
         \begin{overpic}[width=\linewidth, height = 4.5cm]{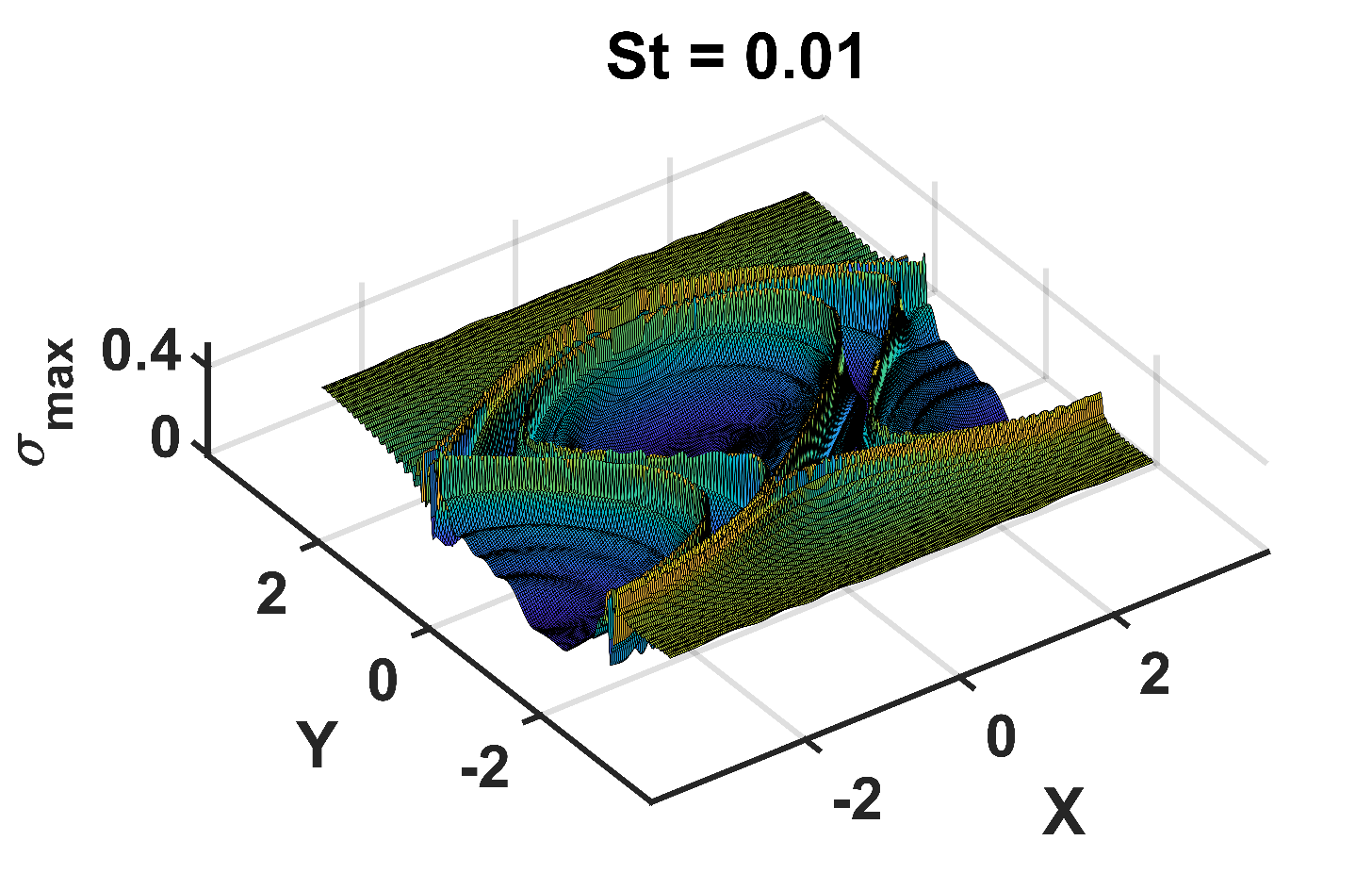}  
        \put(1,75){\textbf{(a)}}
        \end{overpic}   
       
       \label{fig:13a}
        
    \end{subfigure}%
     \begin{subfigure}{0.32\linewidth}
        \centering
        \begin{overpic}[width=\linewidth, height = 4.5cm]{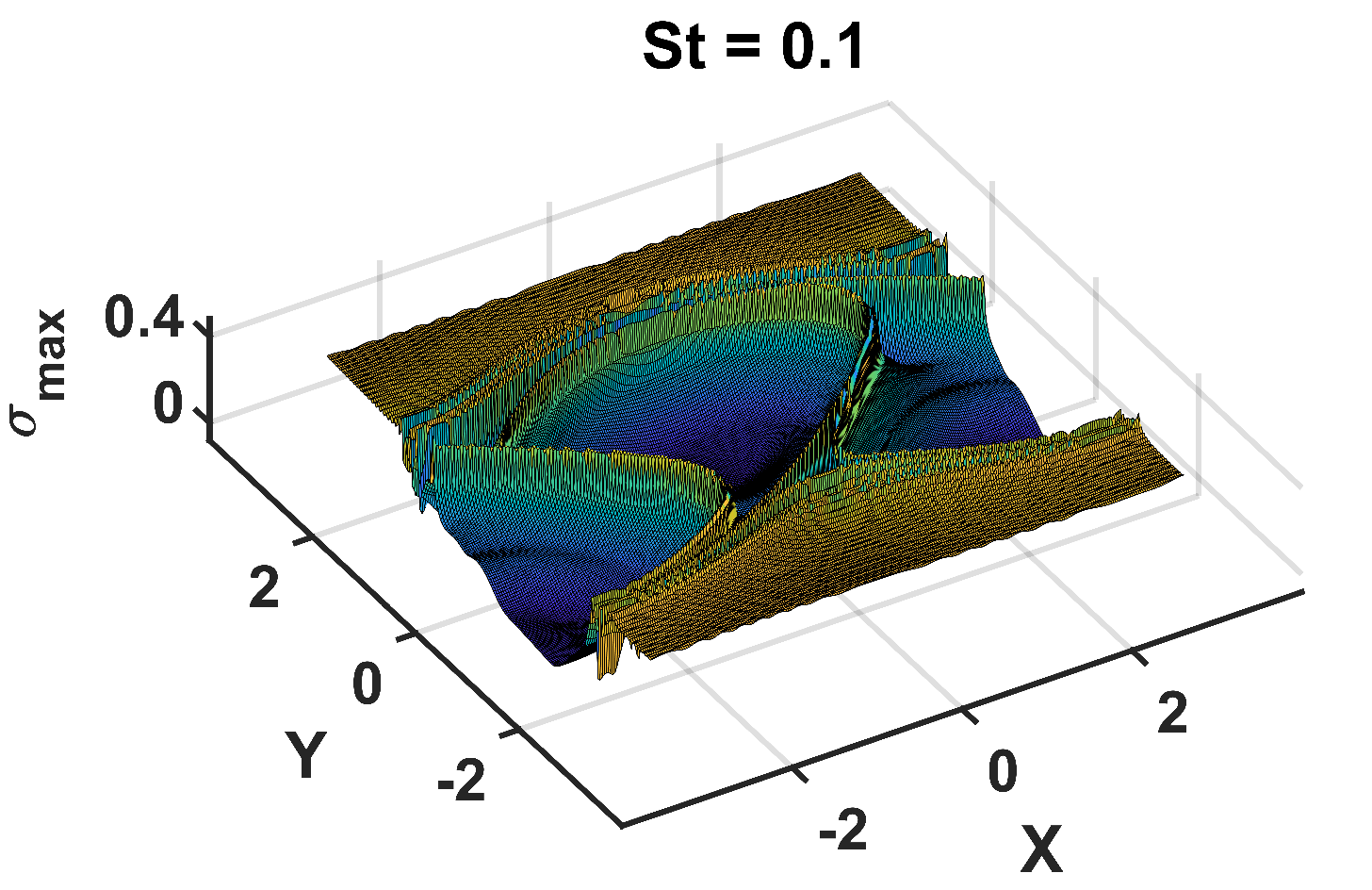}
        \put(1,75){\textbf{(b)}}
        \end{overpic}
           
        \label{fig:13b}
       
    \end{subfigure}%
    \begin{subfigure}{0.32\linewidth}
        \centering
        \begin{overpic}[width=\linewidth, height = 4.5cm]{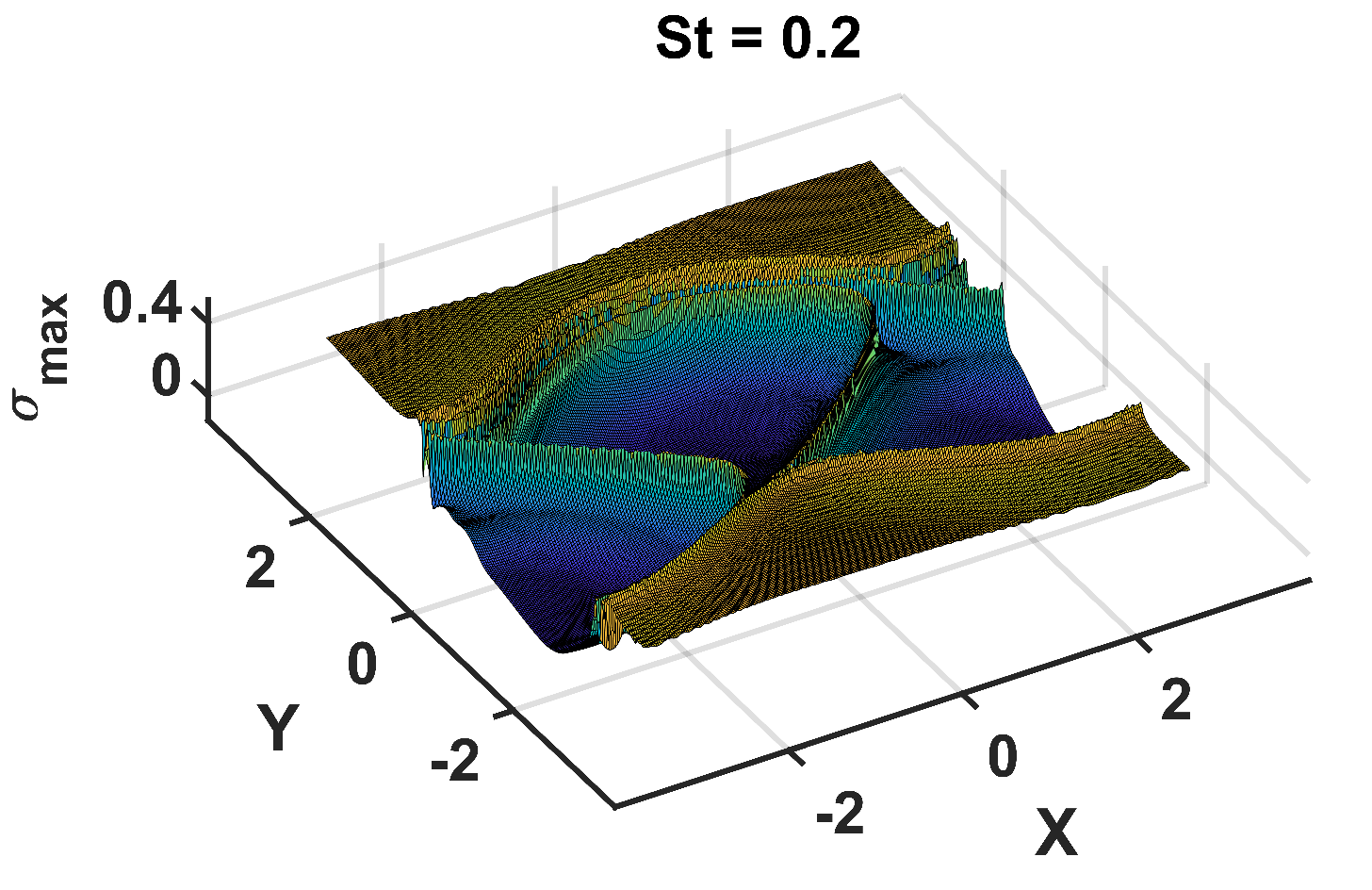} 
        \put(1,75){\textbf{(c)}}
        \end{overpic}
       
       \label{fig:13c}
        
    \end{subfigure}
    \caption{Surface plots for the bubble particles ($R$ = 1) for varying Stokes number say $St$ = 0.01, 0.1, and 0.2 at $t$ = 12 s are plotted. The highest peaks indicate the largest eigenvalues and the shortest peaks indicate the smallest eigenvalues.}
    \label{fig:13}
\end{figure*}

Figure \ref{fig:12} presents a surface plot for aerosol particles, aligning with the results shown in Fig. \ref{fig:10}. The peak points in these surface plots, which we refer to as ``material lines of exponential stretching" in the iFTLE plots, closely match each other. These peak points signify the presence of the largest eigenvalues, while the remaining points indicate slower eigenvalues.

Similarly, Fig. \ref{fig:13} displays a surface plot for bubble particles to corroborate the results presented in Fig. \ref{fig:11}. As observed in Fig. \ref{fig:11}, the bubble particles exhibit fewer ridges, and this is reflected in the surface plots, where fewer peak points are evident.

In conclusion, these surface plots serve as a valuable visual confirmation of the iFTLE results. They clearly highlight regions of significant exponential stretching (represented by peak points with the largest eigenvalues) and regions of slower stretching, reinforcing the insights gained from the iFTLE analysis. This alignment further emphasizes the key role of Stokes number ($St$) in differentiating the mixing behavior of aerosol and bubble particles.

\section{Conclusion}\label{sec: fourth}

In this paper, we studied the dynamics of inertial particles on a base fluid flow modeled by the traveling wave flow. The inertial
particles include both the particles that are less denser than the fluid -  bubbles, and the particles that are more denser than the fluid - aerosols.
The bubbles and aerosols are modeled by a simplified version of the Maxey-Riley equation. 
The dynamics of fluid flow is defined using flow fields wherein measurements
are made at fixed locations in phase space and are referred as Eulerian picture. 
In contrast, the inertial particle transport in fluid flows are depicted
by Lagrangian picture wherein the individual particle trajectories are tracked as they evolve in time.

We computed and visualized the Lagrangian coherent structures of the base fluid flow which are given by the ridges of the finite-time Lyapunov exponent (FTLEs) of the flow. The FTLEs help us to map the region mixing only of the base flow and not to the inertial particle transport.
To capture the inertial particle mixing and transport, we computed the inertial-FTLE (iFTLEs) and visualized the ridges of the iFTLEs. 

In this paper, we have explored the behavior of both the tracer and inertial particles based on the interplay of key parameters: $R$, and $St$, with FTLE and iFTLE fields serving as pivotal tools for our analysis.
By systematically varying these parameters, we analyzed the behavior of aerosol and bubble particles through generating corresponding FTLE fields.
 A significant outcome of our work was the ability to discern whether inertial particles were attracted to or repelled from the attracting manifolds. 
 Particularly noteworthy was the influence of time variation, as inertial particles were advected forward in time, they were gradually drawn towards the attracting manifolds. 

Also, we noted that the density of ridges in the FTLE plots increased with time, indicating the increase in rate of mixing at these ridges.
Our findings also underscored the role of the Stokes number ($St$). As $St$ increased, inertial particles dissipated in phase-space, and corresponding iFTLE contours featured more pronounced ridges. This observation had implications for the segregation of inertial particles based on $St$, as iFTLE values served as indicators of mixing. Notably, aerosol particles exhibited more iFTLE ridges, indicating better mixing, while bubble particles had fewer iFTLE ridges, suggesting lower mixing efficiency. This emphasized that optimal mixing occurs at different Stokes numbers for these two particle types.

To validate our results, we introduced surface plots for the corresponding iFTLE fields, highlighting regions with the highest and lowest peaks, corresponding to the largest and slowest eigenvalues. These surface plots offered an intuitive confirmation of our findings.

This study primarily focused on a two-dimensional, steady, and incompressible flow system. Future directions include extending our work to unsteady systems and investigating the dynamics of inertial particles in three-dimensional flows, reflecting real-world scenarios. Furthermore, there is potential for applying the insights gained in the realm of turbulence control \cite{brunton2015closed}, with the prospect of devising control strategies for segregating particles based on the Stokes number.  Also, FTLEs and iFTLEs can further be employed to investigate the role of various factors, viz. vorticity, acceleration, and strain, that causes preferential concentration of the inertial particles.

\bibliography{apssamp}

\providecommand{\noopsort}[1]{}\providecommand{\singleletter}[1]{#1}%
\begin{thebibliography}{45}%
\makeatletter
\providecommand \@ifxundefined [1]{%
 \@ifx{#1\undefined}
}%
\providecommand \@ifnum [1]{%
 \ifnum #1\expandafter \@firstoftwo
 \else \expandafter \@secondoftwo
 \fi
}%
\providecommand \@ifx [1]{%
 \ifx #1\expandafter \@firstoftwo
 \else \expandafter \@secondoftwo
 \fi
}%
\providecommand \natexlab [1]{#1}%
\providecommand \enquote  [1]{``#1''}%
\providecommand \bibnamefont  [1]{#1}%
\providecommand \bibfnamefont [1]{#1}%
\providecommand \citenamefont [1]{#1}%
\providecommand \href@noop [0]{\@secondoftwo}%
\providecommand \href [0]{\begingroup \@sanitize@url \@href}%
\providecommand \@href[1]{\@@startlink{#1}\@@href}%
\providecommand \@@href[1]{\endgroup#1\@@endlink}%
\providecommand \@sanitize@url [0]{\catcode `\\12\catcode `\$12\catcode
  `\&12\catcode `\#12\catcode `\^12\catcode `\_12\catcode `\%12\relax}%
\providecommand \@@startlink[1]{}%
\providecommand \@@endlink[0]{}%
\providecommand \url  [0]{\begingroup\@sanitize@url \@url }%
\providecommand \@url [1]{\endgroup\@href {#1}{\urlprefix }}%
\providecommand \urlprefix  [0]{URL }%
\providecommand \Eprint [0]{\href }%
\providecommand \doibase [0]{https://doi.org/}%
\providecommand \selectlanguage [0]{\@gobble}%
\providecommand \bibinfo  [0]{\@secondoftwo}%
\providecommand \bibfield  [0]{\@secondoftwo}%
\providecommand \translation [1]{[#1]}%
\providecommand \BibitemOpen [0]{}%
\providecommand \bibitemStop [0]{}%
\providecommand \bibitemNoStop [0]{.\EOS\space}%
\providecommand \EOS [0]{\spacefactor3000\relax}%
\providecommand \BibitemShut  [1]{\csname bibitem#1\endcsname}%
\let\auto@bib@innerbib\@empty
\bibitem [{\citenamefont {Poisson}(1831)}]{poisson}%
  \BibitemOpen
  \bibfield  {author} {\bibinfo {author} {\bibfnamefont {S.~D.}\ \bibnamefont
  {Poisson}},\ }\href@noop {} {\emph {\bibinfo {title} {M{\'e}moire sur les
  {\'e}quations g{\'e}n{\'e}rales de l'{\'e}quilibre et du mouvement des corps
  solides {\'e}lastiques et de fluides}}}\ (\bibinfo  {publisher} {L'imprimerie
  Royale},\ \bibinfo {year} {1831})\BibitemShut {NoStop}%
\bibitem [{\citenamefont {Maxey}\ and\ \citenamefont
  {Riley}(1983)}]{maxey1983equation}%
  \BibitemOpen
  \bibfield  {author} {\bibinfo {author} {\bibfnamefont {M.~R.}\ \bibnamefont
  {Maxey}}\ and\ \bibinfo {author} {\bibfnamefont {J.~J.}\ \bibnamefont
  {Riley}},\ }\bibfield  {title} {\bibinfo {title} {Equation of motion for a
  small rigid sphere in a nonuniform flow},\ }\href@noop {} {\bibfield
  {journal} {\bibinfo  {journal} {The Physics of Fluids}\ }\textbf {\bibinfo
  {volume} {26}},\ \bibinfo {pages} {883} (\bibinfo {year} {1983})}\BibitemShut
  {NoStop}%
\bibitem [{\citenamefont {Bec}\ \emph {et~al.}(2007)\citenamefont {Bec},
  \citenamefont {Biferale}, \citenamefont {Cencini}, \citenamefont {Lanotte},
  \citenamefont {Musacchio},\ and\ \citenamefont {Toschi}}]{bec2007heavy}%
  \BibitemOpen
  \bibfield  {author} {\bibinfo {author} {\bibfnamefont {J.}~\bibnamefont
  {Bec}}, \bibinfo {author} {\bibfnamefont {L.}~\bibnamefont {Biferale}},
  \bibinfo {author} {\bibfnamefont {M.}~\bibnamefont {Cencini}}, \bibinfo
  {author} {\bibfnamefont {A.}~\bibnamefont {Lanotte}}, \bibinfo {author}
  {\bibfnamefont {S.}~\bibnamefont {Musacchio}},\ and\ \bibinfo {author}
  {\bibfnamefont {F.}~\bibnamefont {Toschi}},\ }\bibfield  {title} {\bibinfo
  {title} {Heavy particle concentration in turbulence at dissipative and
  inertial scales},\ }\href@noop {} {\bibfield  {journal} {\bibinfo  {journal}
  {Physical review letters}\ }\textbf {\bibinfo {volume} {98}},\ \bibinfo
  {pages} {084502} (\bibinfo {year} {2007})}\BibitemShut {NoStop}%
\bibitem [{\citenamefont {Squires}\ and\ \citenamefont
  {Eaton}(1991)}]{squires1991preferential}%
  \BibitemOpen
  \bibfield  {author} {\bibinfo {author} {\bibfnamefont {K.~D.}\ \bibnamefont
  {Squires}}\ and\ \bibinfo {author} {\bibfnamefont {J.~K.}\ \bibnamefont
  {Eaton}},\ }\bibfield  {title} {\bibinfo {title} {Preferential concentration
  of particles by turbulence},\ }\href@noop {} {\bibfield  {journal} {\bibinfo
  {journal} {Physics of Fluids A: Fluid Dynamics}\ }\textbf {\bibinfo {volume}
  {3}},\ \bibinfo {pages} {1169} (\bibinfo {year} {1991})}\BibitemShut
  {NoStop}%
\bibitem [{\citenamefont {Riley}\ and\ \citenamefont
  {Patterson~Jr}(1974)}]{riley1974diffusion}%
  \BibitemOpen
  \bibfield  {author} {\bibinfo {author} {\bibfnamefont {J.~J.}\ \bibnamefont
  {Riley}}\ and\ \bibinfo {author} {\bibfnamefont {G.}~\bibnamefont
  {Patterson~Jr}},\ }\bibfield  {title} {\bibinfo {title} {Diffusion
  experiments with numerically integrated isotropic turbulence},\ }\href@noop
  {} {\bibfield  {journal} {\bibinfo  {journal} {The Physics of Fluids}\
  }\textbf {\bibinfo {volume} {17}},\ \bibinfo {pages} {292} (\bibinfo {year}
  {1974})}\BibitemShut {NoStop}%
\bibitem [{\citenamefont {Bec}(2003)}]{bec2003fractal}%
  \BibitemOpen
  \bibfield  {author} {\bibinfo {author} {\bibfnamefont {J.}~\bibnamefont
  {Bec}},\ }\bibfield  {title} {\bibinfo {title} {Fractal clustering of
  inertial particles in random flows},\ }\href@noop {} {\bibfield  {journal}
  {\bibinfo  {journal} {Physics of fluids}\ }\textbf {\bibinfo {volume} {15}},\
  \bibinfo {pages} {L81} (\bibinfo {year} {2003})}\BibitemShut {NoStop}%
\bibitem [{\citenamefont {Bec}\ \emph {et~al.}(2005)\citenamefont {Bec},
  \citenamefont {Celani}, \citenamefont {Cencini},\ and\ \citenamefont
  {Musacchio}}]{bec2005clustering}%
  \BibitemOpen
  \bibfield  {author} {\bibinfo {author} {\bibfnamefont {J.}~\bibnamefont
  {Bec}}, \bibinfo {author} {\bibfnamefont {A.}~\bibnamefont {Celani}},
  \bibinfo {author} {\bibfnamefont {M.}~\bibnamefont {Cencini}},\ and\ \bibinfo
  {author} {\bibfnamefont {S.}~\bibnamefont {Musacchio}},\ }\bibfield  {title}
  {\bibinfo {title} {Clustering and collisions of heavy particles in random
  smooth flows},\ }\href@noop {} {\bibfield  {journal} {\bibinfo  {journal}
  {Physics of Fluids}\ }\textbf {\bibinfo {volume} {17}} (\bibinfo {year}
  {2005})}\BibitemShut {NoStop}%
\bibitem [{\citenamefont {Maxey}\ and\ \citenamefont
  {Corrsin}(1986)}]{maxey1986gravitational}%
  \BibitemOpen
  \bibfield  {author} {\bibinfo {author} {\bibfnamefont {M.~t.}\ \bibnamefont
  {Maxey}}\ and\ \bibinfo {author} {\bibfnamefont {S.}~\bibnamefont
  {Corrsin}},\ }\bibfield  {title} {\bibinfo {title} {Gravitational settling of
  aerosol particles in randomly oriented cellular flow fields},\ }\href@noop {}
  {\bibfield  {journal} {\bibinfo  {journal} {Journal of Atmospheric Sciences}\
  }\textbf {\bibinfo {volume} {43}},\ \bibinfo {pages} {1112} (\bibinfo {year}
  {1986})}\BibitemShut {NoStop}%
\bibitem [{\citenamefont {Maxey}(1987{\natexlab{a}})}]{maxey1987motion}%
  \BibitemOpen
  \bibfield  {author} {\bibinfo {author} {\bibfnamefont {M.}~\bibnamefont
  {Maxey}},\ }\bibfield  {title} {\bibinfo {title} {The motion of small
  spherical particles in a cellular flow field},\ }\href@noop {} {\bibfield
  {journal} {\bibinfo  {journal} {The Physics of fluids}\ }\textbf {\bibinfo
  {volume} {30}},\ \bibinfo {pages} {1915} (\bibinfo {year}
  {1987}{\natexlab{a}})}\BibitemShut {NoStop}%
\bibitem [{\citenamefont {Maxey}(1987{\natexlab{b}})}]{maxey1987gravitational}%
  \BibitemOpen
  \bibfield  {author} {\bibinfo {author} {\bibfnamefont {M.~R.}\ \bibnamefont
  {Maxey}},\ }\bibfield  {title} {\bibinfo {title} {The gravitational settling
  of aerosol particles in homogeneous turbulence and random flow fields},\
  }\href@noop {} {\bibfield  {journal} {\bibinfo  {journal} {Journal of fluid
  mechanics}\ }\textbf {\bibinfo {volume} {174}},\ \bibinfo {pages} {441}
  (\bibinfo {year} {1987}{\natexlab{b}})}\BibitemShut {NoStop}%
\bibitem [{\citenamefont {Rubin}\ \emph {et~al.}(1995)\citenamefont {Rubin},
  \citenamefont {Jones},\ and\ \citenamefont {Maxey}}]{rubin1995settling}%
  \BibitemOpen
  \bibfield  {author} {\bibinfo {author} {\bibfnamefont {J.}~\bibnamefont
  {Rubin}}, \bibinfo {author} {\bibfnamefont {C.}~\bibnamefont {Jones}},\ and\
  \bibinfo {author} {\bibfnamefont {M.}~\bibnamefont {Maxey}},\ }\bibfield
  {title} {\bibinfo {title} {Settling and asymptotic motion of aerosol
  particles in a cellular flow field},\ }\href@noop {} {\bibfield  {journal}
  {\bibinfo  {journal} {Journal of Nonlinear Science}\ }\textbf {\bibinfo
  {volume} {5}},\ \bibinfo {pages} {337} (\bibinfo {year} {1995})}\BibitemShut
  {NoStop}%
\bibitem [{\citenamefont {Shaw}\ \emph {et~al.}(1998)\citenamefont {Shaw},
  \citenamefont {Reade}, \citenamefont {Collins},\ and\ \citenamefont
  {Verlinde}}]{shaw1998preferential}%
  \BibitemOpen
  \bibfield  {author} {\bibinfo {author} {\bibfnamefont {R.~A.}\ \bibnamefont
  {Shaw}}, \bibinfo {author} {\bibfnamefont {W.~C.}\ \bibnamefont {Reade}},
  \bibinfo {author} {\bibfnamefont {L.~R.}\ \bibnamefont {Collins}},\ and\
  \bibinfo {author} {\bibfnamefont {J.}~\bibnamefont {Verlinde}},\ }\bibfield
  {title} {\bibinfo {title} {Preferential concentration of cloud droplets by
  turbulence: Effects on the early evolution of cumulus cloud droplet
  spectra},\ }\href@noop {} {\bibfield  {journal} {\bibinfo  {journal} {Journal
  of the atmospheric sciences}\ }\textbf {\bibinfo {volume} {55}},\ \bibinfo
  {pages} {1965} (\bibinfo {year} {1998})}\BibitemShut {NoStop}%
\bibitem [{\citenamefont {Mezi{\'c}}\ \emph {et~al.}(2010)\citenamefont
  {Mezi{\'c}}, \citenamefont {Loire}, \citenamefont {Fonoberov},\ and\
  \citenamefont {Hogan}}]{mezic2010new}%
  \BibitemOpen
  \bibfield  {author} {\bibinfo {author} {\bibfnamefont {I.}~\bibnamefont
  {Mezi{\'c}}}, \bibinfo {author} {\bibfnamefont {S.}~\bibnamefont {Loire}},
  \bibinfo {author} {\bibfnamefont {V.~A.}\ \bibnamefont {Fonoberov}},\ and\
  \bibinfo {author} {\bibfnamefont {P.}~\bibnamefont {Hogan}},\ }\bibfield
  {title} {\bibinfo {title} {A new mixing diagnostic and gulf oil spill
  movement},\ }\href@noop {} {\bibfield  {journal} {\bibinfo  {journal}
  {Science}\ }\textbf {\bibinfo {volume} {330}},\ \bibinfo {pages} {486}
  (\bibinfo {year} {2010})}\BibitemShut {NoStop}%
\bibitem [{\citenamefont {Tang}\ \emph {et~al.}(2012)\citenamefont {Tang},
  \citenamefont {Knutson}, \citenamefont {Mahalov},\ and\ \citenamefont
  {Dimitrova}}]{tang2012geometry}%
  \BibitemOpen
  \bibfield  {author} {\bibinfo {author} {\bibfnamefont {W.}~\bibnamefont
  {Tang}}, \bibinfo {author} {\bibfnamefont {B.}~\bibnamefont {Knutson}},
  \bibinfo {author} {\bibfnamefont {A.}~\bibnamefont {Mahalov}},\ and\ \bibinfo
  {author} {\bibfnamefont {R.}~\bibnamefont {Dimitrova}},\ }\bibfield  {title}
  {\bibinfo {title} {The geometry of inertial particle mixing in urban flows,
  from deterministic and random displacement models},\ }\href@noop {}
  {\bibfield  {journal} {\bibinfo  {journal} {Physics of Fluids}\ }\textbf
  {\bibinfo {volume} {24}} (\bibinfo {year} {2012})}\BibitemShut {NoStop}%
\bibitem [{\citenamefont {Beron-Vera}\ \emph {et~al.}(2008)\citenamefont
  {Beron-Vera}, \citenamefont {Olascoaga},\ and\ \citenamefont
  {Goni}}]{beron2008oceanic}%
  \BibitemOpen
  \bibfield  {author} {\bibinfo {author} {\bibfnamefont {F.~J.}\ \bibnamefont
  {Beron-Vera}}, \bibinfo {author} {\bibfnamefont {M.~J.}\ \bibnamefont
  {Olascoaga}},\ and\ \bibinfo {author} {\bibfnamefont {G.}~\bibnamefont
  {Goni}},\ }\bibfield  {title} {\bibinfo {title} {Oceanic mesoscale eddies as
  revealed by lagrangian coherent structures},\ }\href@noop {} {\bibfield
  {journal} {\bibinfo  {journal} {Geophysical Research Letters}\ }\textbf
  {\bibinfo {volume} {35}} (\bibinfo {year} {2008})}\BibitemShut {NoStop}%
\bibitem [{\citenamefont {Nencioli}\ \emph {et~al.}(2011)\citenamefont
  {Nencioli}, \citenamefont {d'Ovidio}, \citenamefont {Doglioli},\ and\
  \citenamefont {Petrenko}}]{nencioli2011surface}%
  \BibitemOpen
  \bibfield  {author} {\bibinfo {author} {\bibfnamefont {F.}~\bibnamefont
  {Nencioli}}, \bibinfo {author} {\bibfnamefont {F.}~\bibnamefont {d'Ovidio}},
  \bibinfo {author} {\bibfnamefont {A.}~\bibnamefont {Doglioli}},\ and\
  \bibinfo {author} {\bibfnamefont {A.}~\bibnamefont {Petrenko}},\ }\bibfield
  {title} {\bibinfo {title} {Surface coastal circulation patterns by in-situ
  detection of lagrangian coherent structures},\ }\href@noop {} {\bibfield
  {journal} {\bibinfo  {journal} {Geophysical Research Letters}\ }\textbf
  {\bibinfo {volume} {38}} (\bibinfo {year} {2011})}\BibitemShut {NoStop}%
\bibitem [{\citenamefont {Espinosa-Gayosso}\ \emph {et~al.}(2015)\citenamefont
  {Espinosa-Gayosso}, \citenamefont {Ghisalberti}, \citenamefont {Ivey},\ and\
  \citenamefont {Jones}}]{espinosa2015density}%
  \BibitemOpen
  \bibfield  {author} {\bibinfo {author} {\bibfnamefont {A.}~\bibnamefont
  {Espinosa-Gayosso}}, \bibinfo {author} {\bibfnamefont {M.}~\bibnamefont
  {Ghisalberti}}, \bibinfo {author} {\bibfnamefont {G.~N.}\ \bibnamefont
  {Ivey}},\ and\ \bibinfo {author} {\bibfnamefont {N.~L.}\ \bibnamefont
  {Jones}},\ }\bibfield  {title} {\bibinfo {title} {Density-ratio effects on
  the capture of suspended particles in aquatic systems},\ }\href@noop {}
  {\bibfield  {journal} {\bibinfo  {journal} {Journal of Fluid Mechanics}\
  }\textbf {\bibinfo {volume} {783}},\ \bibinfo {pages} {191} (\bibinfo {year}
  {2015})}\BibitemShut {NoStop}%
\bibitem [{\citenamefont {Peng}\ and\ \citenamefont
  {Dabiri}(2009)}]{peng2009transport}%
  \BibitemOpen
  \bibfield  {author} {\bibinfo {author} {\bibfnamefont {J.}~\bibnamefont
  {Peng}}\ and\ \bibinfo {author} {\bibfnamefont {J.}~\bibnamefont {Dabiri}},\
  }\bibfield  {title} {\bibinfo {title} {Transport of inertial particles by
  lagrangian coherent structures: application to predator--prey interaction in
  jellyfish feeding},\ }\href@noop {} {\bibfield  {journal} {\bibinfo
  {journal} {Journal of Fluid Mechanics}\ }\textbf {\bibinfo {volume} {623}},\
  \bibinfo {pages} {75} (\bibinfo {year} {2009})}\BibitemShut {NoStop}%
\bibitem [{\citenamefont {Sapsis}\ and\ \citenamefont
  {Haller}(2009)}]{sapsis2009inertial}%
  \BibitemOpen
  \bibfield  {author} {\bibinfo {author} {\bibfnamefont {T.}~\bibnamefont
  {Sapsis}}\ and\ \bibinfo {author} {\bibfnamefont {G.}~\bibnamefont
  {Haller}},\ }\bibfield  {title} {\bibinfo {title} {Inertial particle dynamics
  in a hurricane},\ }\href@noop {} {\bibfield  {journal} {\bibinfo  {journal}
  {Journal of the Atmospheric Sciences}\ }\textbf {\bibinfo {volume} {66}},\
  \bibinfo {pages} {2481} (\bibinfo {year} {2009})}\BibitemShut {NoStop}%
\bibitem [{\citenamefont {Sapsis}\ and\ \citenamefont
  {Haller}(2008)}]{sapsis2008instabilities}%
  \BibitemOpen
  \bibfield  {author} {\bibinfo {author} {\bibfnamefont {T.}~\bibnamefont
  {Sapsis}}\ and\ \bibinfo {author} {\bibfnamefont {G.}~\bibnamefont
  {Haller}},\ }\bibfield  {title} {\bibinfo {title} {Instabilities in the
  dynamics of neutrally buoyant particles},\ }\href@noop {} {\bibfield
  {journal} {\bibinfo  {journal} {Physics of fluids}\ }\textbf {\bibinfo
  {volume} {20}} (\bibinfo {year} {2008})}\BibitemShut {NoStop}%
\bibitem [{\citenamefont {Haller}\ and\ \citenamefont
  {Sapsis}(2008)}]{haller2008inertial}%
  \BibitemOpen
  \bibfield  {author} {\bibinfo {author} {\bibfnamefont {G.}~\bibnamefont
  {Haller}}\ and\ \bibinfo {author} {\bibfnamefont {T.}~\bibnamefont
  {Sapsis}},\ }\bibfield  {title} {\bibinfo {title} {Where do inertial
  particles go in fluid flows?},\ }\href@noop {} {\bibfield  {journal}
  {\bibinfo  {journal} {Physica D: Nonlinear Phenomena}\ }\textbf {\bibinfo
  {volume} {237}},\ \bibinfo {pages} {573} (\bibinfo {year}
  {2008})}\BibitemShut {NoStop}%
\bibitem [{\citenamefont {Weiss}(1991)}]{weiss1991transport}%
  \BibitemOpen
  \bibfield  {author} {\bibinfo {author} {\bibfnamefont {J.~B.}\ \bibnamefont
  {Weiss}},\ }\bibfield  {title} {\bibinfo {title} {Transport and mixing in
  traveling waves},\ }\href@noop {} {\bibfield  {journal} {\bibinfo  {journal}
  {Physics of Fluids A: Fluid Dynamics}\ }\textbf {\bibinfo {volume} {3}},\
  \bibinfo {pages} {1379} (\bibinfo {year} {1991})}\BibitemShut {NoStop}%
\bibitem [{\citenamefont {Weiss}\ and\ \citenamefont
  {Knobloch}(1989)}]{weiss1989mass}%
  \BibitemOpen
  \bibfield  {author} {\bibinfo {author} {\bibfnamefont {J.~B.}\ \bibnamefont
  {Weiss}}\ and\ \bibinfo {author} {\bibfnamefont {E.}~\bibnamefont
  {Knobloch}},\ }\bibfield  {title} {\bibinfo {title} {Mass transport and
  mixing by modulated traveling waves},\ }\href@noop {} {\bibfield  {journal}
  {\bibinfo  {journal} {Physical Review A}\ }\textbf {\bibinfo {volume} {40}},\
  \bibinfo {pages} {2579} (\bibinfo {year} {1989})}\BibitemShut {NoStop}%
\bibitem [{\citenamefont {Fung}\ and\ \citenamefont
  {Vassilicos}(1998)}]{fung1998two}%
  \BibitemOpen
  \bibfield  {author} {\bibinfo {author} {\bibfnamefont {J.~C.~H.}\
  \bibnamefont {Fung}}\ and\ \bibinfo {author} {\bibfnamefont {J.~C.}\
  \bibnamefont {Vassilicos}},\ }\bibfield  {title} {\bibinfo {title}
  {Two-particle dispersion in turbulentlike flows},\ }\href@noop {} {\bibfield
  {journal} {\bibinfo  {journal} {Physical Review E}\ }\textbf {\bibinfo
  {volume} {57}},\ \bibinfo {pages} {1677} (\bibinfo {year}
  {1998})}\BibitemShut {NoStop}%
\bibitem [{\citenamefont {Chen}\ \emph {et~al.}(2006)\citenamefont {Chen},
  \citenamefont {Goto},\ and\ \citenamefont {Vassilicos}}]{chen2006turbulent}%
  \BibitemOpen
  \bibfield  {author} {\bibinfo {author} {\bibfnamefont {L.}~\bibnamefont
  {Chen}}, \bibinfo {author} {\bibfnamefont {S.}~\bibnamefont {Goto}},\ and\
  \bibinfo {author} {\bibfnamefont {J.}~\bibnamefont {Vassilicos}},\ }\bibfield
   {title} {\bibinfo {title} {Turbulent clustering of stagnation points and
  inertial particles},\ }\href@noop {} {\bibfield  {journal} {\bibinfo
  {journal} {Journal of Fluid Mechanics}\ }\textbf {\bibinfo {volume} {553}},\
  \bibinfo {pages} {143} (\bibinfo {year} {2006})}\BibitemShut {NoStop}%
\bibitem [{\citenamefont {Biferale}\ \emph {et~al.}(2005)\citenamefont
  {Biferale}, \citenamefont {Boffetta}, \citenamefont {Celani}, \citenamefont
  {Devenish}, \citenamefont {Lanotte},\ and\ \citenamefont
  {Toschi}}]{biferale2005lagrangian}%
  \BibitemOpen
  \bibfield  {author} {\bibinfo {author} {\bibfnamefont {L.}~\bibnamefont
  {Biferale}}, \bibinfo {author} {\bibfnamefont {G.}~\bibnamefont {Boffetta}},
  \bibinfo {author} {\bibfnamefont {A.}~\bibnamefont {Celani}}, \bibinfo
  {author} {\bibfnamefont {B.}~\bibnamefont {Devenish}}, \bibinfo {author}
  {\bibfnamefont {A.}~\bibnamefont {Lanotte}},\ and\ \bibinfo {author}
  {\bibfnamefont {F.}~\bibnamefont {Toschi}},\ }\bibfield  {title} {\bibinfo
  {title} {Lagrangian statistics of particle pairs in homogeneous isotropic
  turbulence},\ }\href@noop {} {\bibfield  {journal} {\bibinfo  {journal}
  {Physics of Fluids}\ }\textbf {\bibinfo {volume} {17}} (\bibinfo {year}
  {2005})}\BibitemShut {NoStop}%
\bibitem [{\citenamefont {Batchelor}(1952)}]{batchelor1952diffusion}%
  \BibitemOpen
  \bibfield  {author} {\bibinfo {author} {\bibfnamefont {G.~K.}\ \bibnamefont
  {Batchelor}},\ }\bibfield  {title} {\bibinfo {title} {Diffusion in a field of
  homogeneous turbulence: Ii. the relative motion of particles},\ }in\
  \href@noop {} {\emph {\bibinfo {booktitle} {Mathematical Proceedings of the
  Cambridge Philosophical Society}}},\ Vol.~\bibinfo {volume} {48}\ (\bibinfo
  {organization} {Cambridge University Press},\ \bibinfo {year} {1952})\ pp.\
  \bibinfo {pages} {345--362}\BibitemShut {NoStop}%
\bibitem [{\citenamefont {Boffetta}\ and\ \citenamefont
  {Sokolov}(2002)}]{boffetta2002statistics}%
  \BibitemOpen
  \bibfield  {author} {\bibinfo {author} {\bibfnamefont {G.}~\bibnamefont
  {Boffetta}}\ and\ \bibinfo {author} {\bibfnamefont {I.~M.}\ \bibnamefont
  {Sokolov}},\ }\bibfield  {title} {\bibinfo {title} {Statistics of
  two-particle dispersion in two-dimensional turbulence},\ }\href@noop {}
  {\bibfield  {journal} {\bibinfo  {journal} {Physics of fluids}\ }\textbf
  {\bibinfo {volume} {14}},\ \bibinfo {pages} {3224} (\bibinfo {year}
  {2002})}\BibitemShut {NoStop}%
\bibitem [{\citenamefont {Tallapragada}\ and\ \citenamefont
  {Ross}(2008)}]{tallapragada2008particle}%
  \BibitemOpen
  \bibfield  {author} {\bibinfo {author} {\bibfnamefont {P.}~\bibnamefont
  {Tallapragada}}\ and\ \bibinfo {author} {\bibfnamefont {S.~D.}\ \bibnamefont
  {Ross}},\ }\bibfield  {title} {\bibinfo {title} {Particle segregation by
  stokes number for small neutrally buoyant spheres in a fluid},\ }\href@noop
  {} {\bibfield  {journal} {\bibinfo  {journal} {Physical Review E}\ }\textbf
  {\bibinfo {volume} {78}},\ \bibinfo {pages} {036308} (\bibinfo {year}
  {2008})}\BibitemShut {NoStop}%
\bibitem [{\citenamefont {Mehlig}\ \emph {et~al.}(2005)\citenamefont {Mehlig},
  \citenamefont {Wilkinson}, \citenamefont {Duncan}, \citenamefont {Weber},\
  and\ \citenamefont {Ljunggren}}]{mehlig2005aggregation}%
  \BibitemOpen
  \bibfield  {author} {\bibinfo {author} {\bibfnamefont {B.}~\bibnamefont
  {Mehlig}}, \bibinfo {author} {\bibfnamefont {M.}~\bibnamefont {Wilkinson}},
  \bibinfo {author} {\bibfnamefont {K.}~\bibnamefont {Duncan}}, \bibinfo
  {author} {\bibfnamefont {T.}~\bibnamefont {Weber}},\ and\ \bibinfo {author}
  {\bibfnamefont {M.}~\bibnamefont {Ljunggren}},\ }\bibfield  {title} {\bibinfo
  {title} {Aggregation of inertial particles in random flows},\ }\href@noop {}
  {\bibfield  {journal} {\bibinfo  {journal} {Physical Review E}\ }\textbf
  {\bibinfo {volume} {72}},\ \bibinfo {pages} {051104} (\bibinfo {year}
  {2005})}\BibitemShut {NoStop}%
\bibitem [{\citenamefont {Tallapragada}\ \emph {et~al.}(2011)\citenamefont
  {Tallapragada}, \citenamefont {Ross},\ and\ \citenamefont
  {Schmale}}]{tallapragada2011lagrangian}%
  \BibitemOpen
  \bibfield  {author} {\bibinfo {author} {\bibfnamefont {P.}~\bibnamefont
  {Tallapragada}}, \bibinfo {author} {\bibfnamefont {S.~D.}\ \bibnamefont
  {Ross}},\ and\ \bibinfo {author} {\bibfnamefont {D.~G.}\ \bibnamefont
  {Schmale}},\ }\bibfield  {title} {\bibinfo {title} {Lagrangian coherent
  structures are associated with fluctuations in airborne microbial
  populations},\ }\href@noop {} {\bibfield  {journal} {\bibinfo  {journal}
  {Chaos: An Interdisciplinary Journal of Nonlinear Science}\ }\textbf
  {\bibinfo {volume} {21}} (\bibinfo {year} {2011})}\BibitemShut {NoStop}%
\bibitem [{\citenamefont {Tang}\ \emph {et~al.}(2009)\citenamefont {Tang},
  \citenamefont {Haller}, \citenamefont {Baik},\ and\ \citenamefont
  {Ryu}}]{tang2009locating}%
  \BibitemOpen
  \bibfield  {author} {\bibinfo {author} {\bibfnamefont {W.}~\bibnamefont
  {Tang}}, \bibinfo {author} {\bibfnamefont {G.}~\bibnamefont {Haller}},
  \bibinfo {author} {\bibfnamefont {J.-J.}\ \bibnamefont {Baik}},\ and\
  \bibinfo {author} {\bibfnamefont {Y.-H.}\ \bibnamefont {Ryu}},\ }\bibfield
  {title} {\bibinfo {title} {Locating an atmospheric contamination source using
  slow manifolds},\ }\href@noop {} {\bibfield  {journal} {\bibinfo  {journal}
  {Physics of Fluids}\ }\textbf {\bibinfo {volume} {21}} (\bibinfo {year}
  {2009})}\BibitemShut {NoStop}%
\bibitem [{\citenamefont {Beron-Vera}\ \emph {et~al.}(2015)\citenamefont
  {Beron-Vera}, \citenamefont {Olascoaga}, \citenamefont {Haller},
  \citenamefont {Farazmand}, \citenamefont {Tri{\~n}anes},\ and\ \citenamefont
  {Wang}}]{beron2015dissipative}%
  \BibitemOpen
  \bibfield  {author} {\bibinfo {author} {\bibfnamefont {F.~J.}\ \bibnamefont
  {Beron-Vera}}, \bibinfo {author} {\bibfnamefont {M.~J.}\ \bibnamefont
  {Olascoaga}}, \bibinfo {author} {\bibfnamefont {G.}~\bibnamefont {Haller}},
  \bibinfo {author} {\bibfnamefont {M.}~\bibnamefont {Farazmand}}, \bibinfo
  {author} {\bibfnamefont {J.}~\bibnamefont {Tri{\~n}anes}},\ and\ \bibinfo
  {author} {\bibfnamefont {Y.}~\bibnamefont {Wang}},\ }\bibfield  {title}
  {\bibinfo {title} {Dissipative inertial transport patterns near coherent
  lagrangian eddies in the ocean},\ }\href@noop {} {\bibfield  {journal}
  {\bibinfo  {journal} {Chaos: An Interdisciplinary Journal of Nonlinear
  Science}\ }\textbf {\bibinfo {volume} {25}} (\bibinfo {year}
  {2015})}\BibitemShut {NoStop}%
\bibitem [{\citenamefont {P{\'e}rez-Munuzuri}(2015)}]{perez2015clustering}%
  \BibitemOpen
  \bibfield  {author} {\bibinfo {author} {\bibfnamefont {V.}~\bibnamefont
  {P{\'e}rez-Munuzuri}},\ }\bibfield  {title} {\bibinfo {title} {Clustering of
  inertial particles in compressible chaotic flows},\ }\href@noop {} {\bibfield
   {journal} {\bibinfo  {journal} {Physical Review E}\ }\textbf {\bibinfo
  {volume} {91}},\ \bibinfo {pages} {052906} (\bibinfo {year}
  {2015})}\BibitemShut {NoStop}%
\bibitem [{\citenamefont {Babiano}\ \emph {et~al.}(2000)\citenamefont
  {Babiano}, \citenamefont {Cartwright}, \citenamefont {Piro},\ and\
  \citenamefont {Provenzale}}]{babiano2000dynamics}%
  \BibitemOpen
  \bibfield  {author} {\bibinfo {author} {\bibfnamefont {A.}~\bibnamefont
  {Babiano}}, \bibinfo {author} {\bibfnamefont {J.~H.}\ \bibnamefont
  {Cartwright}}, \bibinfo {author} {\bibfnamefont {O.}~\bibnamefont {Piro}},\
  and\ \bibinfo {author} {\bibfnamefont {A.}~\bibnamefont {Provenzale}},\
  }\bibfield  {title} {\bibinfo {title} {Dynamics of a small neutrally buoyant
  sphere in a fluid and targeting in hamiltonian systems},\ }\href@noop {}
  {\bibfield  {journal} {\bibinfo  {journal} {Physical Review Letters}\
  }\textbf {\bibinfo {volume} {84}},\ \bibinfo {pages} {5764} (\bibinfo {year}
  {2000})}\BibitemShut {NoStop}%
\bibitem [{\citenamefont {Solomon}\ and\ \citenamefont
  {Gollub}(1988)}]{solomon1988chaotic}%
  \BibitemOpen
  \bibfield  {author} {\bibinfo {author} {\bibfnamefont {T.}~\bibnamefont
  {Solomon}}\ and\ \bibinfo {author} {\bibfnamefont {J.~P.}\ \bibnamefont
  {Gollub}},\ }\bibfield  {title} {\bibinfo {title} {Chaotic particle transport
  in time-dependent rayleigh-b{\'e}nard convection},\ }\href@noop {} {\bibfield
   {journal} {\bibinfo  {journal} {Physical Review A}\ }\textbf {\bibinfo
  {volume} {38}},\ \bibinfo {pages} {6280} (\bibinfo {year}
  {1988})}\BibitemShut {NoStop}%
\bibitem [{\citenamefont {T{\'e}l}\ \emph {et~al.}(2005)\citenamefont
  {T{\'e}l}, \citenamefont {de~Moura}, \citenamefont {Grebogi},\ and\
  \citenamefont {K{\'a}rolyi}}]{tel2005chemical}%
  \BibitemOpen
  \bibfield  {author} {\bibinfo {author} {\bibfnamefont {T.}~\bibnamefont
  {T{\'e}l}}, \bibinfo {author} {\bibfnamefont {A.}~\bibnamefont {de~Moura}},
  \bibinfo {author} {\bibfnamefont {C.}~\bibnamefont {Grebogi}},\ and\ \bibinfo
  {author} {\bibfnamefont {G.}~\bibnamefont {K{\'a}rolyi}},\ }\bibfield
  {title} {\bibinfo {title} {Chemical and biological activity in open flows: A
  dynamical system approach},\ }\href@noop {} {\bibfield  {journal} {\bibinfo
  {journal} {Physics reports}\ }\textbf {\bibinfo {volume} {413}},\ \bibinfo
  {pages} {91} (\bibinfo {year} {2005})}\BibitemShut {NoStop}%
\bibitem [{\citenamefont {Michaelides}(1997)}]{michaelides1997transient}%
  \BibitemOpen
  \bibfield  {author} {\bibinfo {author} {\bibfnamefont {E.~E.}\ \bibnamefont
  {Michaelides}},\ }\href@noop {} {\emph {\bibinfo {title} {The transient
  equation of motion for particles, bubbles, and droplets}}}\ (\bibinfo {year}
  {1997})\BibitemShut {NoStop}%
\bibitem [{\citenamefont {Schuster}(2012)}]{schuster2012transport}%
  \BibitemOpen
  \bibfield  {author} {\bibinfo {author} {\bibfnamefont {H.~G.}\ \bibnamefont
  {Schuster}},\ }\href@noop {} {\emph {\bibinfo {title} {Transport and Mixing
  in Laminar Flows: From Microfluidics to Oceanic Currents}}}\ (\bibinfo
  {publisher} {John Wiley \& Sons},\ \bibinfo {year} {2012})\BibitemShut
  {NoStop}%
\bibitem [{\citenamefont {Haller}(2001)}]{haller2001distinguished}%
  \BibitemOpen
  \bibfield  {author} {\bibinfo {author} {\bibfnamefont {G.}~\bibnamefont
  {Haller}},\ }\bibfield  {title} {\bibinfo {title} {Distinguished material
  surfaces and coherent structures in three-dimensional fluid flows},\
  }\href@noop {} {\bibfield  {journal} {\bibinfo  {journal} {Physica D:
  Nonlinear Phenomena}\ }\textbf {\bibinfo {volume} {149}},\ \bibinfo {pages}
  {248} (\bibinfo {year} {2001})}\BibitemShut {NoStop}%
\bibitem [{\citenamefont {Shadden}\ \emph {et~al.}(2005)\citenamefont
  {Shadden}, \citenamefont {Lekien},\ and\ \citenamefont
  {Marsden}}]{shadden2005definition}%
  \BibitemOpen
  \bibfield  {author} {\bibinfo {author} {\bibfnamefont {S.~C.}\ \bibnamefont
  {Shadden}}, \bibinfo {author} {\bibfnamefont {F.}~\bibnamefont {Lekien}},\
  and\ \bibinfo {author} {\bibfnamefont {J.~E.}\ \bibnamefont {Marsden}},\
  }\bibfield  {title} {\bibinfo {title} {Definition and properties of
  lagrangian coherent structures from finite-time lyapunov exponents in
  two-dimensional aperiodic flows},\ }\href@noop {} {\bibfield  {journal}
  {\bibinfo  {journal} {Physica D: Nonlinear Phenomena}\ }\textbf {\bibinfo
  {volume} {212}},\ \bibinfo {pages} {271} (\bibinfo {year}
  {2005})}\BibitemShut {NoStop}%
\bibitem [{\citenamefont {Knobloch}\ and\ \citenamefont
  {Weiss}(1987)}]{knobloch1987chaotic}%
  \BibitemOpen
  \bibfield  {author} {\bibinfo {author} {\bibfnamefont {E.}~\bibnamefont
  {Knobloch}}\ and\ \bibinfo {author} {\bibfnamefont {J.}~\bibnamefont
  {Weiss}},\ }\bibfield  {title} {\bibinfo {title} {Chaotic advection by
  modulated traveling waves},\ }\href@noop {} {\bibfield  {journal} {\bibinfo
  {journal} {Physical Review A}\ }\textbf {\bibinfo {volume} {36}},\ \bibinfo
  {pages} {1522} (\bibinfo {year} {1987})}\BibitemShut {NoStop}%
\bibitem [{\citenamefont {Brunton}\ and\ \citenamefont
  {Rowley}(2010)}]{brunton2010fast}%
  \BibitemOpen
  \bibfield  {author} {\bibinfo {author} {\bibfnamefont {S.~L.}\ \bibnamefont
  {Brunton}}\ and\ \bibinfo {author} {\bibfnamefont {C.~W.}\ \bibnamefont
  {Rowley}},\ }\bibfield  {title} {\bibinfo {title} {Fast computation of
  finite-time lyapunov exponent fields for unsteady flows},\ }\href@noop {}
  {\bibfield  {journal} {\bibinfo  {journal} {Chaos: An Interdisciplinary
  Journal of Nonlinear Science}\ }\textbf {\bibinfo {volume} {20}} (\bibinfo
  {year} {2010})}\BibitemShut {NoStop}%
\bibitem [{\citenamefont {Sudharsan}\ \emph {et~al.}(2016)\citenamefont
  {Sudharsan}, \citenamefont {Brunton},\ and\ \citenamefont
  {Riley}}]{sudharsan2016lagrangian}%
  \BibitemOpen
  \bibfield  {author} {\bibinfo {author} {\bibfnamefont {M.}~\bibnamefont
  {Sudharsan}}, \bibinfo {author} {\bibfnamefont {S.~L.}\ \bibnamefont
  {Brunton}},\ and\ \bibinfo {author} {\bibfnamefont {J.~J.}\ \bibnamefont
  {Riley}},\ }\bibfield  {title} {\bibinfo {title} {Lagrangian coherent
  structures and inertial particle dynamics},\ }\href@noop {} {\bibfield
  {journal} {\bibinfo  {journal} {Physical Review E}\ }\textbf {\bibinfo
  {volume} {93}},\ \bibinfo {pages} {033108} (\bibinfo {year}
  {2016})}\BibitemShut {NoStop}%
\bibitem [{\citenamefont {Brunton}\ and\ \citenamefont
  {Noack}(2015)}]{brunton2015closed}%
  \BibitemOpen
  \bibfield  {author} {\bibinfo {author} {\bibfnamefont {S.~L.}\ \bibnamefont
  {Brunton}}\ and\ \bibinfo {author} {\bibfnamefont {B.~R.}\ \bibnamefont
  {Noack}},\ }\bibfield  {title} {\bibinfo {title} {Closed-loop turbulence
  control: Progress and challenges},\ }\href@noop {} {\bibfield  {journal}
  {\bibinfo  {journal} {Applied Mechanics Reviews}\ }\textbf {\bibinfo {volume}
  {67}},\ \bibinfo {pages} {050801} (\bibinfo {year} {2015})}\BibitemShut
  {NoStop}%
\end{thebibliography}%

\end{document}